\documentclass[amsmath,amssymb,twocolumn,superscriptaddress,floatfix,preprintnumbers, prb]{revtex4}
\usepackage[pdftex]{graphicx}
\usepackage{subfigure}
\usepackage{dcolumn}
\usepackage{bm}
\usepackage{hyperref}
\usepackage{color}

\newcommand{\bra}[1]{\langle #1 \vert}
\newcommand{\ket}[1]{\vert #1 \rangle}
\newcommand{\braket}[1]{\langle #1 \rangle}

\newcommand{\ud}{\uparrow\downarrow}
\newcommand{\du}{\downarrow\uparrow}

\bibpunct{(}{)}{;}{s}{,}{,}

\begin{document}

\title{Probing unconventional superconductivity in inversion symmetric doped Weyl semimetal}

\author{Youngseok Kim}
\affiliation{Department of Electrical and Computer Engineering, University of Illinois, Urbana, Il, 61801}
\author{Moon Jip Park}
\affiliation{Department of Physics, University of Illinois, Urbana, Illinois 61801, USA}
\author{Matthew J. Gilbert}
\affiliation{Department of Electrical and Computer Engineering, University of Illinois, Urbana, Il, 61801}
\affiliation{Micro and Nanotechnology Laboratory, University of Illinois, Urbana, Illinois 61801, USA}

\date{\today}

\begin{abstract}
Unconventional superconductivity has been predicted to arise in the topologically non-trivial Fermi surface of doped inversion symmetric Weyl semimetals (WSM). In particular, Fulde-Ferrell-Larkin-Ovchinnikov (FFLO) and nodal BCS states are theoretically predicted to be possible superconductor pairing states in inversion symmetric doped WSM. In an effort to resolve preferred pairing state, we theoretically study two separate four terminal quantum transport methods that each exhibit a unique electrical signature in the presence of FFLO and nodal BCS states in doped WSMs. We first introduce a Josephson junction that consists of a doped WSM and an s-wave superconductor in which we show that the application of a transverse uniform current in s-wave superconductor effectively cancels the momentum carried by FFLO states in doped WSM. From our numerical analysis, we find a peak in Josephson current amplitude at finite uniform current in s-wave superconductor that serves as an indicator of FFLO states in doped WSMs. Furthermore, we show using a four terminal measurement configuration that the nodal points may be shifted by an application of transverse uniform current in doped WSM. We analyze the topological phase transitions induced by nodal pair annihilation in non-equilibrium by constructing the phase diagram and we find a characteristic decrease in the density of states that serves as a signature of the quantum critical point in the topological phase transition, thereby identifying nodal BCS states in doped WSM.
\end{abstract}
\maketitle

\section{Introduction} 
Rapid progress in the field of topological phases of matter has extended the scope of our understanding from fully gapped insulator to gapless semimetals\cite{Murakami2007, Savrasov2011, Schnyder2014}. An example of which is the Weyl semimetal (WSM), whose low energy excitations are described by three-dimensional Weyl fermions\cite{Murakami2007, Savrasov2011}. The WSM is characterized by its non-degenerate band crossing points referred to as Weyl nodes, where the valence and conduction band touch. Weyl nodes are monopoles of the Berry curvature in momentum space\cite{Fang2003, Murakami2007} and the Fermi surface (FS) enclosing the Weyl node is topologically non-trivial as it carries monopole charge (or Chern number). 
Weyl nodes with opposite monopole charge appear in pairs in the lattice\cite{Nielsen1981_1,Nielsen1981_2} and the pairs of Weyl nodes are responsible for emergent phenomena such as Fermi arcs\cite{Savrasov2011,Murakami2014,Xu2015} and unconventional electromagnetic responses such as negative magneto-resistance and chiral magnetic effect\cite{Hosur2013}. 

The unique physics of WSM motivates further research on one of the most striking differences between semimetals and insulators; the \emph{intrinsic} superconducting phases in doped semimetal. 
Unconventional superconductivity has been shown to arise from the interplay between topologically non-trivial states and superconducting phases of doped WSM\cite{Cho2012, Vivek2014, Burkov2015, Wang2015, Qi2014}.
Specifically, as FS enclosing Weyl nodes must appear in even number\cite{Nielsen1981_1,Nielsen1981_2},
doped WSM facilitates two types of possible superconducting pairings: inter-node and intra-node pairing.
When Weyl nodes with opposite monopole charge are mapped to each other by inversion symmetry, the inter-node pairing exhibits nodal BCS pairing state whose electrical structure is in a close analogy with the $^3$He-A phase\cite{Cho2012, Sato2015, Haldane2015}. On the other hand, the intra-node pairing forms finite momentum carrying superconducting states\cite{Cho2012} known as the Fulde-Ferrell-Larkin-Ovchinnikov (FFLO) states\cite{Ferrell1964, Ovchinnikov1965}. 
While both types of superconducting states are possible, different analysis methods yield different energetically preferred pairing states\cite{Cho2012, Vivek2014, Burkov2015, Wang2015}. 
Assuming even parity pairing (singlet) states in low-energy chiral basis, mean-field calculations show that FFLO pairing is favored\cite{Cho2012}. On the contrary, when one considers odd parity pairing (triplet), a short- and long-range attractive interaction results in FFLO and BCS pairing states as ground states, respectively\cite{Vivek2014}. In the weak-coupling regime, BCS states are energetically preferred, however, FFLO states may have lower energy in the absence of both inversion and time-reversal symmetry, due to the fact that FFLO states rely on low-energy chiral symmetry while electrons in the BCS states are connected either by inversion or time-reversal symmetry\cite{Burkov2015}. 

Although finding energetically preferred pairing is crucial to clarify microscopic details of the superconductivity, it is unclear how to determine a pairing scheme for a given doped WSM. In this regards, we propose a quantum transport method to elucidate the pairing states in doped WSM. More precisely, we focus our discussion on inversion symmetric doped WSM and on two possible unconventional superconducting states: FFLO and nodal BCS states. 
To identify two seemingly distinct superconducting states, we propose two complementary transport methods. In section \ref{sec:FFLO}, we introduce a Josephson junction comprised of a doped WSM and a conventional s-wave superconductor in weak coupling regime to resolve the FFLO states. 
We find that the Josephson current is averaged out to be vanishingly small due to the spatially oscillating order parameter of FFLO states. By driving transverse supercurrent in s-wave superconductor, we show that non-equilibrium s-wave pairing states mimic FFLO states and the Josephson current is restored at finite transverse current, which serves as a signature for FFLO pairing in doped WSM. In section \ref{sec:BCS}, we introduce a system consists of a doped WSM attached with four terminal contacts to identify nodal BCS states. We show that nodal points are shifted in momentum space by tuning transverse DC current, which may result in an annihilation of nodal points and a subsequent topological phase transition. At the critical point of the topological phase transition, we find a distinct peak in longitudinal differential conductance ($dI/dV$) curve inside the superconducting gap that serves as a signature of the nodal BCS states in doped WSM. In section \ref{sec:Conclusion}, we summarize our results and conclude.

\section{Probing FFLO pairing states} \label{sec:FFLO} 
\subsection{System description} \label{sec:sys}
\begin{figure}[t!]  
  \centering
   \includegraphics[width=0.5\textwidth]{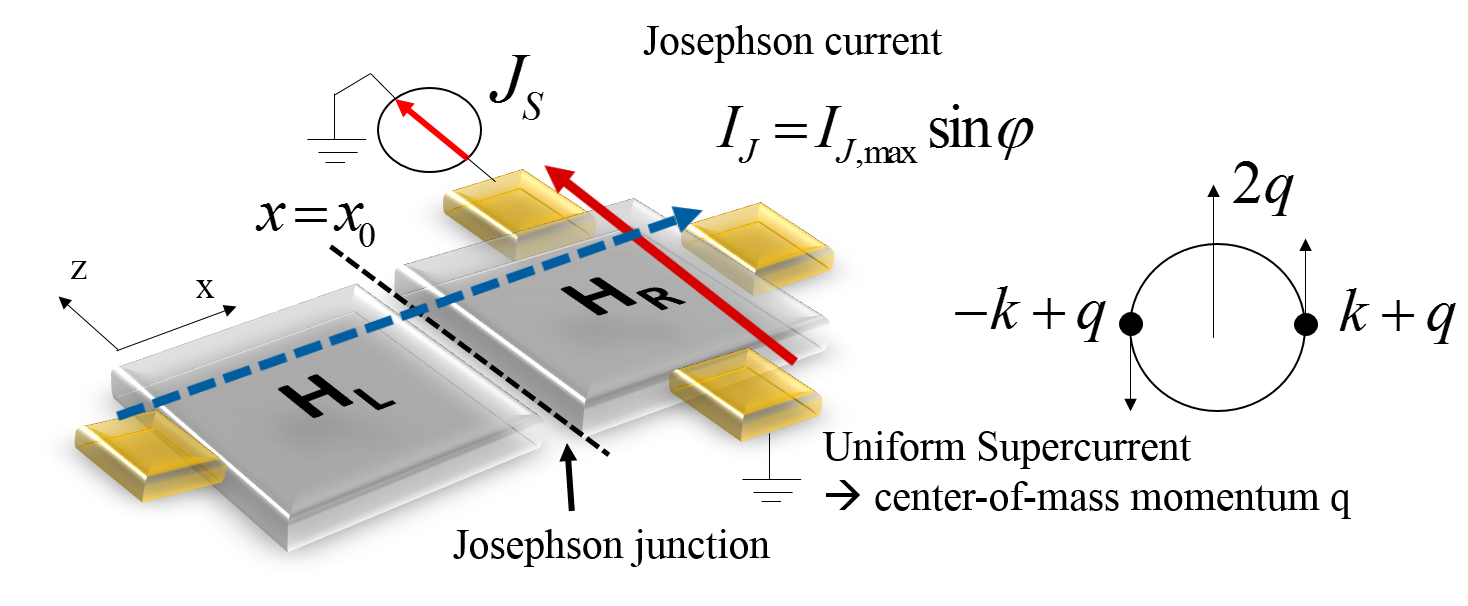}
  \caption{A schematic of the system. $H_L$ is a WSM and $H_R$ is an ordinary metal superconductor. A weak coupling between $H_L$ and $H_R$ is assumed. A Josephson current flows in $\hat{x}$ direction (blue dashed arrow) and a uniform supercurrent in $\hat{z}$ direction (red solid arrow) gives center-of-momentum $q$ to the $H_R$ system. }\label{fig:system}
\end{figure}   
In Fig. (\ref{fig:system}), we consider a Josephson junction that consists of a doped WSM ($H_L$) weakly coupled with a conventional s-wave superconductor ($H_R$). When the system is in the superconducting regime, a Josphson current flows in longitudinal ($\hat{x}$) direction, as shown by the blue dashed arrow in Fig. (\ref{fig:system}), across the junction located at $x=x_0$. The doped inversion symmetric WSM system in this work has two Weyl nodes located at $\pm\mathbf{Q}$ in momentum space. 
Assuming inter-node pairing, a Cooper pair that shares a FS with momenta $\pm \mathbf{Q}+\mathbf{k}$ and $\pm \mathbf{Q}-\mathbf{k}$ forms an FFLO state\cite{Cho2012}.
Therefore, a net momentum of $\pm2\mathbf{Q}$ is carried by the pairing states and the order parameter of the FFLO states has a form $\Psi_{L}(\mathbf{r})=\psi_L (e^{i 2\mathbf{Q}\cdot \mathbf{r}}+e^{-i 2\mathbf{Q}\cdot \mathbf{r}})$ in real space, where $\psi_L$ is an amplitude of the order parameter\cite{Ferrell1964, Ovchinnikov1965}. Assuming uniform BCS pairing for the s-wave superconductor, the superconducting order parameter is $\Psi_R(\mathbf{r})=\psi_R$ and the total Josephson current may be determined as\cite{Yang2000}
\begin{equation} \label{eq:IJ}
I_J\propto \text{Im} \left[
\psi_L\psi_R \int d^2 \mathbf{r} e^{i(2\mathbf{Q}\cdot \mathbf{r}+\delta\varphi)}+e^{i(-2\mathbf{Q}\cdot \mathbf{r}+\delta\varphi)}
\right],
\end{equation}
where $\delta\varphi$ is relative phase difference of two superconducting systems, and the integral covers the entire interface of the Josephson junction. 
In Eq. (\ref{eq:IJ}), $I_J$ vanishes as one integrates over $\mathbf{r}$ due to the spatially oscillating FFLO state order parameter. 
However, previous work\cite{Yang2000} shows that one may effectively cancel the finite momentum $\mathbf{Q}$ by introducing external magnetic field and, as a result, the Josephson current is restored. Although the non-zero Josephson current under applied magnetic field can be utilized to identify FFLO states, the same proposal may not be applicable in the WSM. In the presence of a magnetic field, the low energy Hamiltonian of WSM leaves only 1D chiral mode in the lowest Landau level\cite{Carbotte2013} and, therefore, the intra-node coupling cannot occur. To overcome this situation, we show that a driven supercurrent plays the role of the magnetic field.

In the presence of a uniform supercurrent of s-wave superconductor, as depicted in Fig. (\ref{fig:system}) by the red solid arrow, a Cooper pair aquires a finite center-of-mass momentum $\mathbf{q}$. Then electrons at $\mathbf{k}+\mathbf{q}$ and $-\mathbf{k}+\mathbf{q}$ constitute a Cooper pair with a net momentum of $2\mathbf{q}$. As a result, the s-wave pairing states under non-equilibrium effectively mimic finite-momentum carrying FFLO states with the order parameter\cite{Gennes1966, Tinkham1996} $\Psi_R=\psi_R e^{i2\mathbf{q}\cdot \mathbf{r}}$. Especially, when the momentum $\mathbf{q}$ is parallel to and in a resonance with $\mathbf{Q}$ carried by the FFLO states, the Josephson junction has a non-vanishing $I_J$, which may serve as a signature of FFLO states in doped WSM.
In above scenario, a uniform transverse current, $\mathbf{J}_S$, is carried by Cooper pairs with finite net momentum $2\mathbf{q}$, as indicated by the red solid arrow in Fig. (\ref{fig:system}). $\mathbf{J}_S$ increases linearly as a function of $\mathbf{q}$ both in the conventional s-wave\cite{Bardeen1962} and unconventional nodal superconductor\cite{Maki2004} until $\mathbf{J}_S$ reaches a critical current, or the superconducting phase becomes unstable. In this paper, however, we assume that $\mathbf{J}_S$ is small compared to the critical current, therefore, the supercurrent is proportional to $\mathbf{q}$ (see supplementary material for the calculation of $J_S$ as a function of $q$). Therefore, we utilize $\mathbf{q}$ as a key parameter to describe non-equilibrium states of the superconductor system and plot our main results as a function of $\mathbf{q}$ instead of $\mathbf{J}_S$.

We begin by considering a model lattice Hamiltonian
\begin{equation} \label{eq:Htot}
H=H_L + H_R + H_T,
\end{equation}
where $H_L$ is a doped WSM system and $H_R$ is a metallic s-wave system as depicted in Fig. (\ref{fig:system}). We assume both of the systems are in superconducting phase and they are weakly coupled by a tunneling Hamiltonian, $H_T$. 
We discretize the system in longitudinal ($\hat{x}$) direction in order to consider a Josephson junction at $x=x_0$ with the tunneling Hamiltonian 
\begin{equation} \label{eq:Ht}
H_{T}=\sum_{\mathbf{k},\mathbf{p}} t_{\mathbf{k},\mathbf{p}} (c^\dagger_{\mathbf{k}}(x_0) c_{\mathbf{p}}(x_0) + h.c.),
\end{equation}
where $c^\dagger_{\mathbf{k}}$ is electron creation operator of system $H_L$, $c_{\mathbf{p}}$ is annihilation operator of system $H_R$, $t_{\mathbf{k},\mathbf{p}}$ is a tunneling constant, and $\mathbf{k}$, $\mathbf{p}=(k_y,k_z)$ are momentum of transverse directions. Here, we assume that the tunneling constant is non-zero only at the interface ($x=x_0$). 

For the doped WSM system, we choose a model Hamiltonian which breaks time reversal symmetry but preserves inversion symmetry. Near the Weyl node, we consider a minimal low-energy two-band model of the WSM\cite{Ran2011}
\begin{equation} \label{eq:Hw}
\begin{split}
H_w=&\sum_\mathbf{k}\Big[ \left(
M-2\sum_{\alpha=x,y,z} t_\alpha \cos k_\alpha
\right) \sigma_z  \\
& +2\lambda \left(\sin k_x \sigma_x + \sin k_y \sigma_y\right)-\mu_L\mathbb{I}\Big],
\end{split}
\end{equation}
where $\sigma_{x,y,z}$ are the Pauli matrices for spin, $\mathbb{I}$ is the identity matrix, $\lambda$ is a hopping term in $k_x-k_y$ plane, and $\mu_L$ is the chemical potential in the WSM. In this work, we use a lattice constant of $a=1$ and set $\hbar=1$.
In Eq. (\ref{eq:Hw}), $t_{\alpha=x,y,z}$ is a mass term which determines the position of the Weyl nodes in momentum space. The time-reversal breaking mass term $M=2t_x+2t_y+m$ separates Weyl nodes in the system and we set $m=2t_z\cos Q$ so that two Weyl nodes are located at $\pm\mathbf{Q}=(0,0,\pm Q)$ along the $z$ axis with opposite monopole charge. 
Assuming FFLO pairing, we consider an attractive Hubbard type interaction. The mean-field approximation for the interaction Hamiltonian is
\begin{equation} \label{eq:Hint}
H_{FFLO}=\sum_\mathbf{k}[\Delta_{L1} c^\dagger_{\mathbf{k},\uparrow}c^\dagger_{-\mathbf{k}+2\mathbf{Q},\downarrow}+\Delta_{L2} c^\dagger_{\mathbf{k},\uparrow}c^\dagger_{-\mathbf{k}-2\mathbf{Q},\downarrow}+h.c.].
\end{equation}
where the first (second) term couples electrons in FS enclosing the Weyl node located at $k_z=+Q$ ($-Q$) with a uniform pairing potential $\Delta_{L1}$ ($\Delta_{L2}$).
To see the finite size effect of the junction, we discretize the Hamiltonian in transverse ($\hat{z}$) direction. 
Therefore, the Hamiltonian of Eqs. (\ref{eq:Hw}) and (\ref{eq:Hint}) is discretized in transverse ($\hat{z}$) and longitudinal ($\hat{x}$) direction in real space. As a result, the Bogoliubov-de Gennes (BdG) Hamiltonian is
\begin{equation} \label{eq:HLz}
\begin{split}
&H_L =
\sum_{\mathbf{r},k_y}\mathbf{\Phi}_{r,k_y}^\dagger
\begin{pmatrix}
\tilde{H}_{w}(k_y) & \tilde{H}_{FFLO}(\mathbf{r}) \\
\tilde{H}_{FFLO}^\dagger(\mathbf{r}) &  -\tilde{H}^*_{w}(-k_y)
\end{pmatrix}
\mathbf{\Phi}_{r,k_y}   \\
&+\sum_{\mathbf{r},\alpha,k_y} \left[
\mathbf{\Phi}_{r,k_y}^\dagger
\begin{pmatrix}
\tilde{H}_{w,\alpha} & 0 \\
0 &  -\tilde{H}^*_{w,\alpha}
\end{pmatrix}
\mathbf{\Phi}_{r+\alpha,k_y}
+\text{h.c.}
\right],
\end{split}
\end{equation}
where $\mathbf{\Phi}_{r,k_y}=[c_{\mathbf{r},k_y,\uparrow},c_{\mathbf{r},k_y,\downarrow},c^\dagger_{\mathbf{r},-k_y,\uparrow},c^\dagger_{\mathbf{r},-k_y,\downarrow}]^{T}$, $\mathbf{r}=(x,z)$, and $\alpha=\delta x, \delta z$. The individual components of discretized Hamiltonian are 
\begin{equation}  \label{eq:HLz2}
\begin{split}
&\tilde{H}_{w}(k_y)=[M-2t_y\cos(k_y a)]\sigma_z+2\lambda\sin(k_y a)\sigma_y-\mu_L\mathbb{I}, \\
&\tilde{H}_{w,\delta x}=-i\lambda \sigma_x-t_x\sigma_z,\; \tilde{H}_{w,\delta z}=-t_z\sigma_z, \\
&\tilde{H}_{FFLO}(\mathbf{r})=2\Delta_L \cos(2Qz) i\sigma_y,
\end{split}
\end{equation}
where $\tilde{H}_{w,\delta x}$ and $\tilde{H}_{w,\delta z}$ are the nearest neighbor hopping Hamiltonian in the $\hat{x}$ and $\hat{z}$ direction, respectively, and $\tilde{H}_{FFLO}(\mathbf{r})$ is the superconducting interaction Hamiltonian Fourier transformed to real space. Note that we assume identical pairing potential for each FS, $\Delta_{L1}=\Delta_{L2}=\Delta_{L}$, but following arguments are valid regardless of this assumption.

With the Weyl Hamiltonian defined, we consider a normal metal Hamiltonian defined as
\begin{equation} \label{eq:Hm}
H_m= \sum_{\mathbf{k}}( -t_m(\cos k_x +\cos k_y + \cos k_z)-\mu_R )\mathbb{I},
\end{equation}
where $t_m$ is a hopping term and $\mu_R$ is the chemical potential. In our system, the Cooper pairs in the BCS superconductor aquire $\mathbf{q}=q\hat{z}$ through the application of a uniform supercurrent\cite{Gennes1966, Tinkham1996} in transverse ($\hat{z}$) direction, as indicated in red solid arrow in Fig. (\ref{fig:system}). Then the mean-field approximation to the interaction Hamiltonian is
\begin{equation}
H_{BCS}=\sum_\mathbf{k}[\Delta_{R} c^\dagger_{\mathbf{k}+\mathbf{q}\uparrow}c^\dagger_{-\mathbf{k}+\mathbf{q}\downarrow}+h.c.],
\end{equation}
where $\Delta_R$ is a uniform BCS pairing potential. 
The BdG Hamiltonian is constructed for $H_R$ in a similar manner to Eq. (\ref{eq:HLz}) and discretized in the transverse ($\hat{z}$) and longitudinal ($\hat{x}$) directions. 
Consequently,  
\begin{equation} \label{eq:HRzq}
\begin{split}
&H_R(q) = \sum_{\mathbf{r},k_y}
\mathbf{\Phi}_{r,k_y}^\dagger
\begin{pmatrix}
\tilde{H}_{m}(k_y) & \tilde{H}_{BCS}(\mathbf{r},q) \\
\tilde{H}_{BCS}^\dagger(\mathbf{r},q) &  -\tilde{H}^*_{m}(-k_y)
\end{pmatrix}
\mathbf{\Phi}_{r,k_y}  \\
&+\sum_{\mathbf{r},\alpha,k_y} \left[
\mathbf{\Phi}_{r,k_y}^\dagger
\begin{pmatrix}
\tilde{H}_{m,\alpha} & 0 \\
0 &  -\tilde{H}^*_{m,\alpha}
\end{pmatrix}
\mathbf{\Phi}_{r+\alpha,k_y}
+\text{h.c.}
\right],
\end{split}
\end{equation}
where the discretized Hamiltonians are
\begin{equation}
\begin{split}
&\tilde{H}_{m}(k_y) = (-t_m\cos k_z -\mu_R)\mathbb{I}, \\
&\tilde{H}_{m,\delta x} = -(t_m/2)\mathbb{I},\; \tilde{H}_{m,\delta z} = -(t_m/2)\mathbb{I}, \\
&\tilde{H}_{BCS}(\mathbf{r},q) = \Delta_R e^{i2qz} i\sigma_y. \\
\end{split}
\end{equation}
Here, $\tilde{H}_{m,\delta x}$ and $\tilde{H}_{m,\delta z}$ are the nearest neighbor hopping Hamiltonian and $\tilde{H}_{BCS}(\mathbf{r},q)$ is the interaction Hamiltonian Fourier transformed to real space.

\subsection{Josephson current}
Having defined lattice Hamiltonian for $H_{L/R}$, we may calculate the Josephson current between the doped WSM and s-wave superconductor. Assuming a weak coupling limit, the tunneling Hamiltonian $H_T$ in Eq. (\ref{eq:Ht}) can be treated as a perturbation. From the Ginzburg-Landau theory, we may determine the Josephson current\cite{Yang2000}
\begin{equation} \label{eq:IJ2}
I_J=\text{Im} \left[
t_c\int d\mathbf{r}_\parallel^2 
\Psi^\dagger_{BCS}(\mathbf{r}_\parallel)\Psi_{FFLO}(\mathbf{r}_\parallel) 
\right],
\end{equation}
where $t_c$ is a coupling constant, $\Psi_{BCS}$ and $\Psi_{FFLO}$ are order parameters of s-wave superconductor and doped WSM system, respectively. The integration in Eq. (\ref{eq:IJ2}) is performed over the interface of the Jospehson junction $\mathbf{r}_\parallel=(x_0,y,z)$, whose longitudinal ($\hat{x}$) direction is fixed at the junction position $x=x_0$.
Once we put two superconductors together, the order parameters may differ in phase by $\delta\varphi=\varphi_L-\varphi_R$. Taking account the phase difference, the order parameters in Eq. (\ref{eq:IJ2}) are rewritten as $\Psi_{FFLO}=\Psi_L(\mathbf{r}_\parallel) e^{i\varphi_L}$ and $\Psi_{BCS}=\Psi_R(\mathbf{r}_\parallel,q)e^{i\varphi_R}$, where $\Psi_{L}$ and $\Psi_{R}$ are the order parameters of doped WSM and s-wave superconductor, respectively. Note that the order parameters $\Psi_{L}$ and $\Psi_{R}$ are calculated in \emph{isolated} system as the tunneling Hamiltonian is treated perturbatively. Then, Eq. (\ref{eq:IJ2}) is rewritten as
\begin{equation} \label{eq:IJ3}
\begin{split}
I_J=&\text{Im} \left[
t_c\int d^2\mathbf{r}_\parallel 
\Psi_R^\dagger(\mathbf{r}_\parallel,q)\Psi_L(\mathbf{r}_\parallel)e^{i\delta\varphi}
\right] \\
=&\text{Im} \left[ I_{J,max}(q) e^{i\varphi(q)}e^{i\delta\varphi}  \right] \\
=&I_{J,max}(q) \sin(\varphi(q)+\delta\varphi),
\end{split}
\end{equation}
where $I_{J,max}$ and $\varphi(q)+\delta\varphi$ are the amplitude and phase of the Josephson current, $I_J$. We immediately notice that the Josephson current amplitude, $I_{J,max}$, is a function of momentum $q$. As it is shown in Eq. (\ref{eq:HLz2}), the interaction Hamiltonian of doped WSM oscillates spatially which manifests as a spatial oscillation in the order parameter $\Psi_L$. As a result, $I_{J,max}$ is spatially averaged out and its magnitude vanishes for a sufficiently wide interface ($\gg 1/Q$) at $q=0$. 
The situation, however, may be different when a Cooper pair in s-wave superconductor acquires center-of-mass momentum $q$ by a driven current. The order parameter $\Psi_R$ effectively mimics FFLO states with non-zero momentum $q$ to cancel out the relative spatial variation and, at $q=\pm Q$, $I_{J,max}$ is restored. To evaluate $I_{J,max}$, we take a Fourier transform of both order parameters $\Psi_{L/R}$ in $\hat{y}$ direction
\begin{equation} \label{eq:IJmax}
\begin{split}
&I_{J,max}(q)=
\left\vert t_c\int d^2 \mathbf{r}_\parallel \Psi_R^\dagger(\mathbf{r}_\parallel,q)\Psi_L(\mathbf{r}_\parallel) \right\vert \\
=&
\left\vert t_c\int dz 
\int \frac{dk_y}{2\pi}   
\Psi_R^\dagger(\mathbf{r}_0,k_y,q) 
\Psi_L(\mathbf{r}_0,k_y) \right\vert,
\end{split}
\end{equation}
where $\mathbf{r}_0=(x_0,z)$.
Then the Hamiltonians in Eqs. (\ref{eq:HLz}) and (\ref{eq:HRzq}) are diagonalized and the order parameters $\Psi_L(\mathbf{r},k)=\braket{c_{\uparrow,\mathbf{r},k}c_{\downarrow,\mathbf{r},-k}}_L$ and $\Psi_R(\mathbf{r},k,q)=\braket{c_{\uparrow,\mathbf{r},k}c_{\downarrow,\mathbf{r},-k}}_R$ are evaluated (see appendix \ref{app:OP}). 
In Fig. (\ref{fig:Imax}), we plot $I_{J,max}$ calculated from Eq. (\ref{eq:IJmax}). We see a clear peak in $I_{J,max}$ at $q=\pm Q$ where the momentum $q$ in BCS superconductor cancels the momentum $Q$ carried by FFLO states in WSM.
The oscillations in $I_{J,max}$ are due to the finite size of the lattice having an insufficient sampling of k-space. The width of the peak is decreased as we increase the resolution of the momentum space by increasing the system size. The peak is ideally a delta function at $q=\pm Q$ if the junction size is large enough to satisfy $\Delta k=2\pi/L_z\ll Q$. In the presence of weak disorder, the peak may be shifted as disorder renormalizes mass term of WSM Hamiltonian\cite{Hughes2016}, but persist as the FFLO states discussed here is robust to impurity scattering\cite{Cho2012}. Therefore, the Josephson current amplitude at non-zero transverse ($\hat{z}$) current ($q\neq 0$) may serve as a signature of FFLO states for inversion symmetric doped WSM.

\begin{figure}[t!]  
  \centering
   \includegraphics[width=0.5\textwidth]{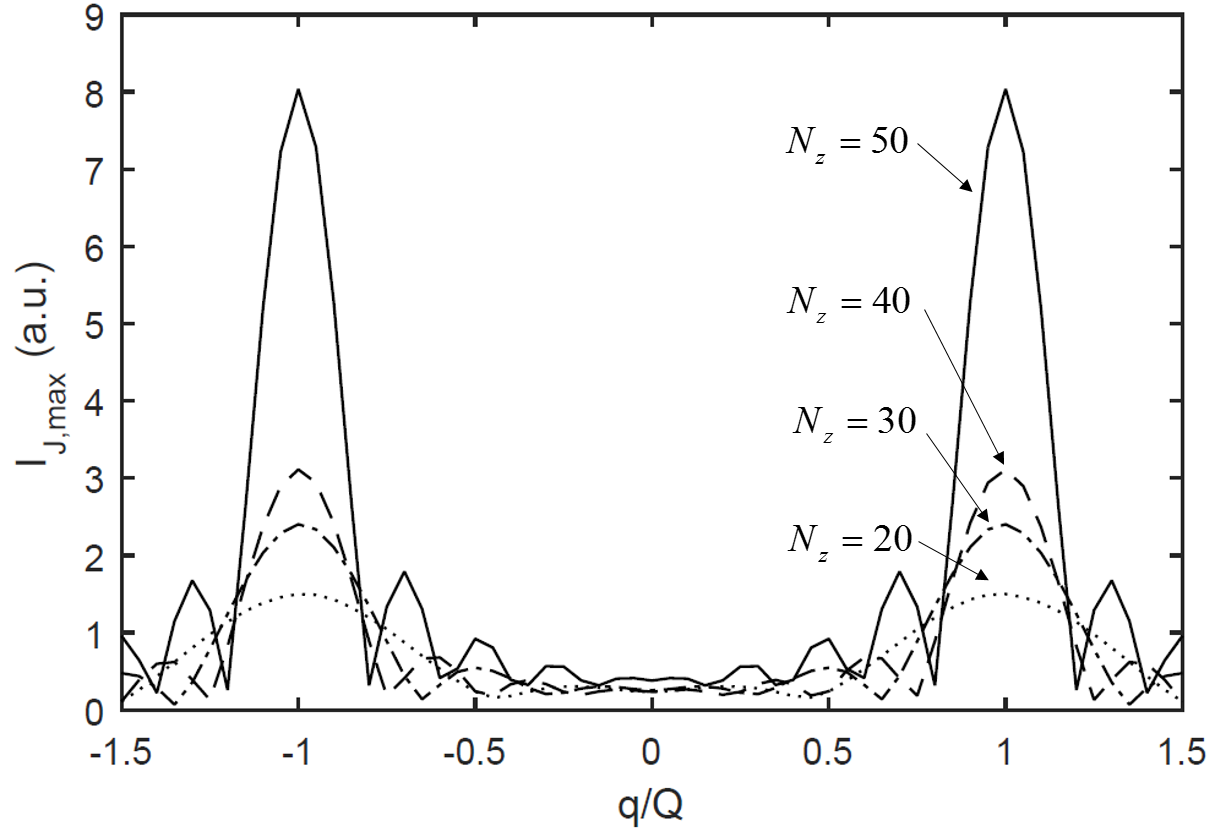}
  \caption{Plot of Josephson current maximum $I_{J,max}$ in Eq. (\ref{eq:IJmax}) as a function of momentum, $q$, in the BCS superconductor as described by $H_R$. There are two clear peaks when the $q$ matches with the $\pm Q$ in doped WSM described by $H_L$. The parameters $t_m=1, \mu_R=0$ are used for $H_R$ and $t_x=0.5,\; t_y=0.5,\; t_z=1.0, \; \lambda=0.5,\; \mu_L/t=0.2$, and $Q=0.1\pi$ are used for $H_L$. The pairing potentials $\Delta_L/t=\Delta_R/t=0.2$ are used and the number of points along the longitudinal direction ($\hat{x}$), $N_x=10$, is fixed for both $H_L$ and $H_R$. In order to see the finite size effect of the Josephson junction, we plot $N_z=20$ to $N_z=50$. }\label{fig:Imax}
\end{figure}   

\section{Probing nodal superconductivity} \label{sec:BCS} 
While intra-node superconducting states are identified by quantum transport signatures in the Josephson junction, applying the same method may not confirm inter-node superconducting states as nodal BCS states do not carry finite momentum and the current response simply returns to conventional Josephson junction results. Instead, we exploit nodal structures of inversion symmetric doped WSM\cite{Cho2012, Balants2012, Sato2015, Haldane2015} and propose a separate quantum transport method to identify nodal BCS superconductivity using a four terminal measurement.

\subsection{Nodal BCS states in doped WSM} \label{sec:BCSA}
As the only prerequisite for nodal superconductivity in doped WSM is the presence of inversion symmetry\cite{Haldane2015}, the inter-node pairing results in nodal superconductivity even in the presence of a uniform BCS pairing potential. Each nodal point carries topologically non-trivial \emph{vorticity} inherited from the monopole charge of the corresponding FS in the normal phase\cite{Haldane2015}. Therefore, each nodal point exhibits similar physics with that of the WSM such as Fermi arcs\cite{Sato2015,Haldane2015,Wang2015}. In addition, the nodal BCS superconductivity fascillitates a zero energy flat band dispersion at its surface that is protected by mirror symmetry\cite{Cho2012, Sato2015,Haldane2015}. 
The flat band zero energy can be experimentally confirmed by zero bias conductance peak at the surface\cite{Sato2015} and may serve as an evidence of nodal superconductivity. However, seeking the zero bias peak may be a difficult task due to the gapless bulk conducting channels\cite{Franz2016}. Instead, we propose to utilize a induced topological phase transition by application of current through the superconducting system.
Here, we show that the nodal points, initially assumed to be well separated in equilibrium, are shifted in momentum space by a uniform supercurrent. Then nodal pair annihilation may occur and the subsequent phase transition depletes available bulk states within the superconducting gap. As a result, the phase transition is captured by a distinct dip in the density of states (DOS or $dI/dV$) and an observation of the dip in non-equilibrium may serve as a signature of nodal BCS superconductivity in doped WSM. 

\subsection{Nodal pair annihilation and energy spectrum}
\begin{figure}[t!]  
  \centering
   \includegraphics[width=0.5\textwidth]{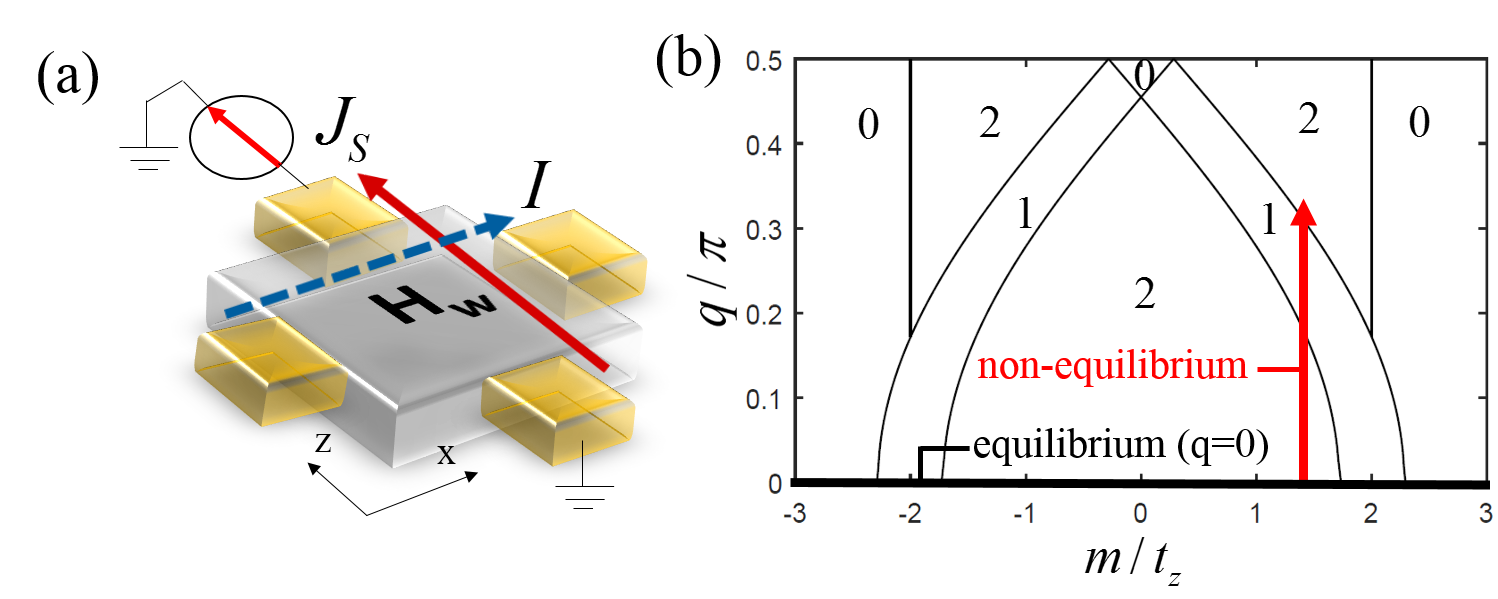}
  \caption{(a) A schematic of the system. Uniform supercurrent $J_S$ is driven to the Weyl superconductor system $H_w$. Differential conductance is read from a current measured in perpendicular direction ($I$). (b) Phase diagram of the number of nodal point pairs from Hamiltonian Eq. (\ref{eq:HBdGw}) at $k_x=k_y=0$. A DOS is obtained in particular direction indicated in red vertical arrow and plotted in Fig. (\ref{fig:DOS}). For WSM Hamiltonian, the same parameters used in Fig. (\ref{fig:Imax}) are adopted. The range of $q$ presented here is $0\leq q\leq\pi/2$ due to the fact that a relevant range of total cooper pair momentum is $\vert2q\vert\leq\pi$.}\label{fig:BCS}
\end{figure}   
To examine the induced topological transition, we assume a four terminal device setup outlined in Fig. (\ref{fig:BCS}a). The red solid arrow in Fig. (\ref{fig:BCS}a) represents a uniform supercurrent driven by external current source, which induces a net momentum shift of Cooper pairs by a momentum $q$ in transverse ($\hat{z}$) direction. In the following argument, we show that the momentum $q$ shifts nodal points in momentum space to induce topological phase transition. To observe the corresponding topological phase transition, we utilize the DOS by measuring a differential conductance in longitudinal ($\hat{x}$) direction shown as a blue dashed arrow in Fig. (\ref{fig:BCS}a).
For inversion symmetric doped WSM, we use the lattice WSM Hamiltonian $H_w=\sum_\mathbf{k} \tilde{H}_w(\mathbf{k})$ in Eq. (\ref{eq:Hw}) with shifted center-of-mass frame by $q$ to account for uniform supercurrent. Assuming uniform BCS pairing, the BdG Hamiltonian is 
\begin{equation} \label{eq:HBdGw}
H_{\text{BdG}}=\sum_{\mathbf{k}, q}\Phi_{\mathbf{k},q}^\dagger
\begin{pmatrix}
\tilde{H}_w(\mathbf{k}+q) &  \tilde{H}_{BCS} \\
\tilde{H}^\dagger_{BCS} & -\tilde{H}^*_w(-\mathbf{k}+q) \\
\end{pmatrix}\Phi_{\mathbf{k},q},
\end{equation}
where $\Phi_{\mathbf{k},q}=[c_{\mathbf{k}+q,\uparrow},c_{\mathbf{k}+q,\downarrow},c^\dagger_{-\mathbf{k}+q,\uparrow},c^\dagger_{-\mathbf{k}+q,\downarrow}]$. In this shifted center-of-mass frame, the mean-field interaction Hamiltonian is defined as $\tilde{H}_{BCS}=\Delta_0 i\sigma_y$, where $\Delta_0$ is a uniform pairing potential.
The position of the nodal points in Eq. (\ref{eq:HBdGw}) is determined by considering the quasi-particle spectrum along the $k_z$ axis. For illustrative purposes, we analyze the Hamiltonian in Eq. (\ref{eq:Hw}) along the $k_z$ direction, which is $\tilde{H}_w(k_z)=[m-2t_z\cos(k_z)]\sigma_z-\mu\mathbb{I}$, by setting $k_x=k_y=0$. Then, Eq. (\ref{eq:HBdGw}) may be rewritten in a block diagonal form as $\begin{pmatrix} \tilde{H}_{\ud} & 0 \\ 0 & \tilde{H}_{\du} \end{pmatrix}$, whose bases in each block are $[c_{\mathbf{k}+q,\uparrow}, c^\dagger_{-\mathbf{k}+q,\downarrow}]$ and $[c_{\mathbf{k}+q,\downarrow},c^\dagger_{-\mathbf{k}+q,\uparrow}]$, respectively. The quasi-particle spectrum along $k_z$ axis is
\begin{equation} \label{eq:Eud}
\begin{split}
E_{\ud}^{\pm}(k_z,q)=&(m-2t_z\cos k_z\cos q) \\
&\pm\sqrt{\Delta_0^2+(\mu-2t_z\sin k_z\sin q)^2}, \\
E_{\du}^{\pm}(k_z,q)=&(2t_z\cos k_z\cos q-m) \\
&\pm\sqrt{\Delta_0^2+(\mu+2t_z\sin k_z\sin q)^2}. \\
\end{split}
\end{equation}
In Eq. (\ref{eq:Eud}), the nodal points are found at the crossings of the quasi-particle spectra.
To see the nodal point dependency on $q$, we further simplify Eq. (\ref{eq:Eud}) by assuming $\mu=0$ and setting the mass term to be $m=2t_z\cos Q$ to place Weyl nodes at $k_z=\pm Q$. We then expand quasi-particle spectrum around $\pm Q$. Specifically, we set $k_z=\pm Q+\delta k_z$ where $\delta k_z \ll Q$ is an infinitesimal deviation from a location of Weyl node in normal phase. Assuming a small $q$ ($q\ll Q$) we obtain,
\begin{equation} \label{eq:Edk}
\begin{split}
E_+(\delta k_z,q)\simeq&[t_z'\delta k_z \pm\sqrt{\Delta_0^2+{t_z'}^2 q^2}] \sigma_z, \\
E_-(\delta k_z,q)\simeq&[-t_z'\delta k_z \pm\sqrt{\Delta_0^2+{t_z'}^2 q^2}] \sigma_z, \\
\end{split}
\end{equation} 
where $E_\pm$ is the quasi-particle spectrum in the vicinity of $k_z=\pm Q$. In Eq. (\ref{eq:Edk}), we set $t_z'=2t_z\sin Q$ and $\sigma_z$ is the Pauli matrix in pseudospin space whose components consist of linear combinations of the eigenfunctions in Eq. (\ref{eq:Eud}). 
Eq. (\ref{eq:Edk}) shows that each FS has two nodal points at $\delta k_z=\pm\sqrt{(\Delta_0/t_z')^2+q^2}$ and the nodal points are shifted as a function of $q$ toward $k_z=0$ and $\pi$.
Due to the particle-hole symmetry, we know that a nodal point pair exists at $(k_z,E)=(k_0,E_0)$ and $(-k_0,-E_0)$, and the pair consists of opposite vorticity by inversion symmetry of WSM. Therefore, by manipulating $q$, the nodal pair with opposite vorticity may be shifted to be annihilated at $k_z=0$ or $\pm\pi$ and the total number of nodal point pairs given by the band topology at equilibrium can be tuned.
In Fig. (\ref{fig:BCS}b), the phase diagram of the system that contains different number of nodal point pairs is shown as a function of the mass term $m$ and momentum $q$, which determines the position of nodal points in equlibrium and non-equlibrium, respectively. The wavevector $q$ is controlled by applied current and $m$ is determined by 
the magnetic order of material or magnetized impurities. 
When $q=0$, the system contains two nodal point pairs for $\vert m\vert \leq2t_z-\sqrt{\Delta_0^2+\mu^2}$. If the normal phase of WSM has Weyl node separation smaller than $2Q\leq \sqrt{\Delta_0^2+\mu^2}$ in momentum space, a pair of nodal points is annihilated as one turns on the superconductivity and, as a result, only one nodal point pair remains in the system. When $\vert m\vert \geq2t_z+\sqrt{\Delta_0^2+\mu^2}$, the system is fully gapped and no nodal point pairs exist. Departing from equilibrium, nodal points are shifted and annihilated by increasing $q$, for example, as shown in the red vertical arrow in Fig. (\ref{fig:BCS}b). 
Note that we only consider a phase diagram when $(k_x,k_y)=(0,0)$. The same arguments are also valid for other high-symmetry points in Brillouin zone such as $(k_x,k_y)=(\pm\pi,\pm\pi)$ which simply replaces mass term $m\rightarrow m+4t_x$ for $(\pm\pi,0)$, $m+4t_y$ for $(0,\pm\pi)$, and $m+4t_x+4t_y$ for $(\pm\pi,\pm\pi)$. Nonetheless, the resulting physics is identical.

\subsection{Signatures of the phase transition}

\begin{figure}[t!]  
  \centering
   \includegraphics[width=0.5\textwidth]{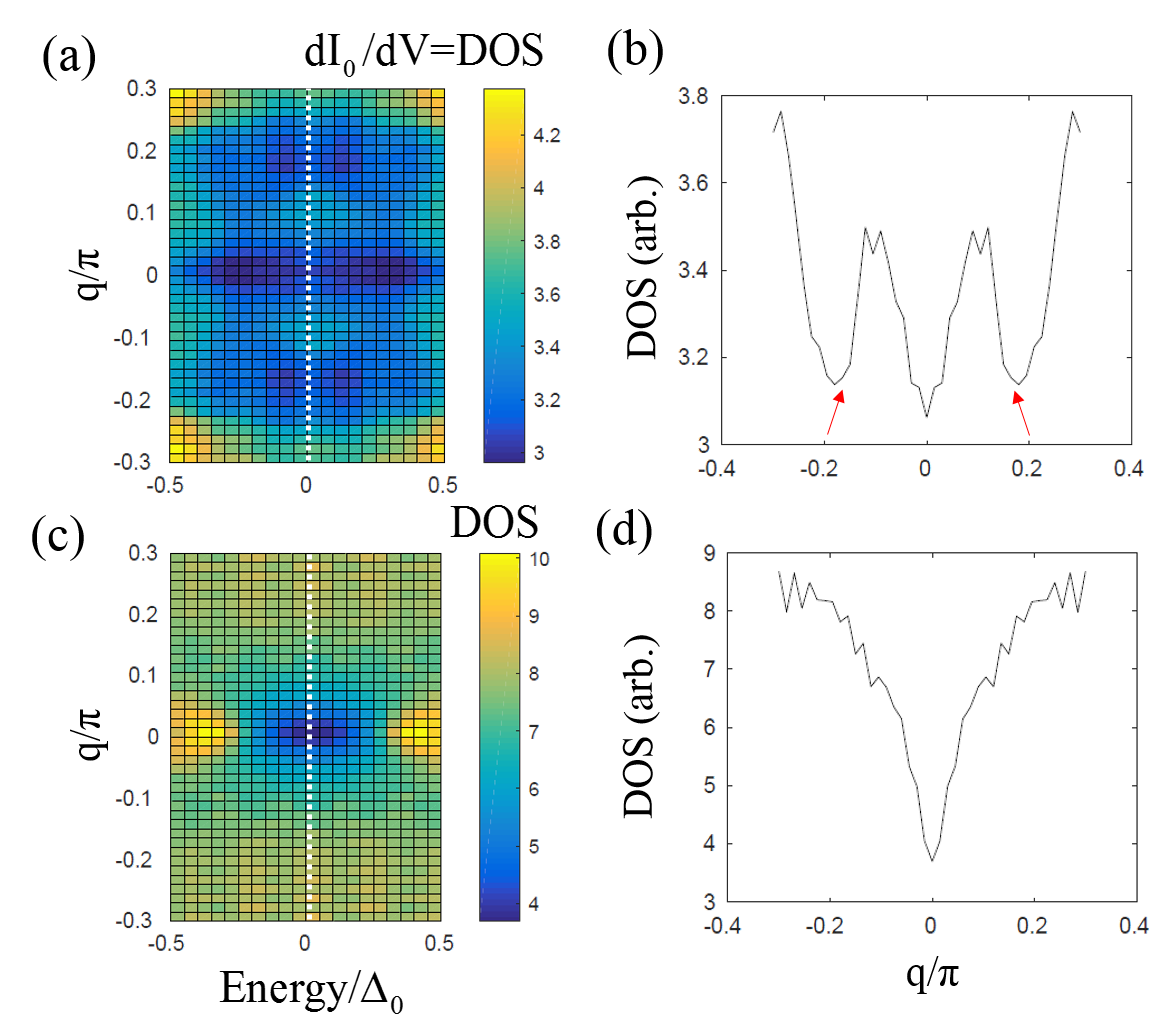}
  \caption{A DOS plot along the vertical red arrow in Fig. (\ref{fig:BCS} b). (a) DOS is plotted as a function of $q$ within a superconducting gap. Pairing potential is set to be $\Delta_0/t_z=0.2$ with (a) $N_y=50$ and (c) $N_y=5$. A chemical potential is $\mu/t_z=0.2$. (b) DOS plot at $E=0$ as a function of $q$ for $N_y=50$ (white dotted line in (a)). (d) DOS plot at $E=0$ as a function of $q$ for $N_y=5$ (white dotted line in (c)).}\label{fig:DOS}
\end{figure}   

When the phase transition occurs under non-equilibrium conditions, the annihilated nodal pairs no longer provide available states within the superconducting gap. As a result, the induced topological transition is observed in the DOS (or $dI/dV$). To examine this, we compute DOS(E,q) as a function of energy and momentum $q$ using the system Green's function\cite{Datta2005}. Note that the system boundary in $y$ direction is open in real space so that we may observe finite-size effects and the surface states contribution. 
To examine the induced topological phase transition and the corresponding DOS, we sweep $q$ at an arbitrary cut of the phase diagram at $m/t_z=2\cos(Q=0.2\pi)\simeq 1.6$. As the red arrow in Fig. (\ref{fig:BCS}b) shows, the phase transition occurs around $q/\pi \simeq 0.1$. Fig. (\ref{fig:DOS}a) shows the corresponding DOS where we set the thickness to $N_y=50$ to avoid finite size effects. Along the horizontal axis at $q=0$ in Fig. (\ref{fig:DOS}a), equilibrium DOS increases quadratically in energy $(\propto E^2)$ within the superconducting gap due to the presence of bulk nodal points, whereas surface states result in non-zero DOS near $E=0$. 
When the system is not in equilibrium ($q\neq 0$), the eigenstates initially separated by a superconducting gap are shifted by $q$ and added to the available low energy states\cite{Maki2004}. As a result, DOS increases as a function of $q$. However, there are distinct drops in magnitude of DOS at certain $q$ as it is seen by following vertical axis in Fig. (\ref{fig:DOS}a). With this particular choice of mass ($m$) in the phase diagram, a pair of nodal points with opposite vorticity moves toward $k_z=0$ and is annihilated at $q/\pi \simeq 0.1$. Further increase in $q$ from this point gaps out the spectrum at $k_z=0$ and a topological phase transition occurs leaving only one pair of nodal point pair in the system. Thus, the available states within the superconducting gap is decreased and the consequent change in nodal structure manifests itself as a dip in the DOS. The dip is clearly observed in the zero energy cut indicated with red arrows in Fig. (\ref{fig:DOS}b). Therefore, the distinct dip of the DOS in non-equilibrium is a signature of quantum critical point which can only occur due to the topological phase transition of the nodal superconductor. 
Note that above arguments are valid for a system where the bulk nodal points are well defined so that their annihilation can be clearly identified. If the bulk nodal points are gapped out by the finite size effect, the signature may not be obvious in the DOS. Fig. (\ref{fig:DOS}c) shows a DOS for a thickness of $N_y=5$ where the bulk states are gapped out by the finite size effect. The DOS within the finite size induced gap is suppressed but finite due to an infinitesimal broadening we introduced in Green's function calculation\cite{Datta2005} and surface states with hybridization gap $E/\Delta\simeq 0.5$. Consequently, in Fig. (\ref{fig:DOS}d), we observe a monotonic increase of DOS as a function of $q$ and no clear signature of nodal point annihilation is observed.

\section{Summary and Conclusion} \label{sec:Conclusion}  
In summary, we study two complementary quantum transport methods to probe FFLO and nodal BCS states in superconducting phase of the inversion symmetric doped WSM. To identify FFLO states, we consider a Josephson junction consisting of a doped WSM and conventional s-wave superconductor. When the junction is in the weak coupling limit, the Josephson current is calculated from the order parameters in lattice Hamiltonian using Ginzburg-Landau theory. The order parameter of the doped WSM oscillates spatially due to the finite momentum, $Q$, carried by FFLO states that results a vanising Josephson current. By driving a uniform current in conventional s-wave superconductor, the order parameter of s-wave superconductor effectively mimics FFLO states carrying a net momentum $q$. When the modulated order parameter effectively cancels $Q$ at $q=\pm Q$, a finite Josephson current is restored. Therefore, the peak in Josephson current in non-equilibrium serves as a direct signature of the presence of FFLO states in doped WSM. 

Additionally, we show that protected nodal points in equilibrium may be shifted by using four contacts quantum transport geometry on doped WSM. The system may undergo an induced topological transition by annihilating the nodal point pairs, which is signalled by an abrupt changes in the DOS (or differential conductance). Using lattice model and Green's function, we observe a distinct dip in DOS as one across a boundary of the phase diagram where a nodal point pair annihilation occurs. Thus, the induced topological phase transition and corresponding singatures in the DOS at non-equilibrium may serve as an indication of the nodal superconductivity in doped WSM.

\begin{acknowledgements} 
M.J.P. and M.J.G. acknowledge financial support
from the Office of Naval Research (ONR) under grant
N0014-11-1-0123 and the National Science Foundation
(NSF) under grant CAREER EECS-1351871. M.J.P. and Y. K. acknowledge useful discussions from Timothy Philip and Yuxuan Wang. M.J.P. acknowledges fruitful discussions from David ChangMo Yang and Gil Young Cho.
\end{acknowledgements} 

\appendix
\section{Order parameter calculation} \label{app:OP}
In this appendix, we summarize the method utilized to obtain order parameter in Eq. (\ref{eq:IJmax}) from BdG Hamiltonian. The Hamiltonians in Eq. (\ref{eq:HLz}) and (\ref{eq:HRzq}) are discretized in $r=(x,z)$ direction with a momentum $k$ in $\hat{y}$ direction. Then the Hamiltonian can be diagonalized from following Bogoliubov transform\cite{Gennes1966}
\begin{equation} \label{eq:tr1}
\begin{split}
\begin{pmatrix}
c_{r,k,\sigma} \\ c_{r,-k,\bar\sigma}^\dagger
\end{pmatrix}&=
\sum_n 
\begin{pmatrix}
u_{n,r,k} & -v_{n,r,k}^* \\
v_{n,r,k} & u_{n,r,k}^* \\
\end{pmatrix}
\begin{pmatrix}
\gamma_{\alpha,n,r,k} \\ \gamma_{\beta,n,r,k}^\dagger
\end{pmatrix} \\
&=
\sum_n 
R_{n,k,z}
\begin{pmatrix}
\gamma_{\alpha,n,k,z} \\ \gamma_{\beta,n,k,z}^\dagger
\end{pmatrix},
\end{split}
\end{equation}
where $\sigma=\uparrow,\downarrow$ is spin index and $\bar\sigma$ stands for an opposite spin with $\sigma$ and a quasi-particle operator index $(\alpha,\beta)=(1,2)$ for $\sigma=\uparrow$ and $(3,4)$ for $\sigma=\downarrow$ for each eigenstate index $n$. Here, we define a basis rotation matrix $R_{n,r,k}$ which diagonalizes the Hamiltonian
\begin{equation} \label{eq:HRRE}
H(r,k)R_{n,r,k}=R_{n,r,k}
\begin{pmatrix}
E_{n,r,k} & 0 \\ 0 & -E_{n,r,k}
\end{pmatrix}.
\end{equation}  
Therefore, we obtain the rotation matrix $R_{n,k,z}$ and corresponding eigenvalue $E_{n,r,k}$ by diagonalizing the Hamiltonian in Eqs. (\ref{eq:HLz}) and (\ref{eq:HRzq}).
Then, an order parameter with uniform s-wave pairing potential is defined as
\begin{equation} \label{eq:OP0}
\Psi(r,k)=
\braket{c_{r,k,\uparrow}c_{r,-k,\downarrow}}.
\end{equation}
The quasi-particle operator $\gamma$ satisfies commutation relation  $\gamma_{\alpha,n}^\dagger\gamma_{\alpha',m}+\gamma_{\alpha',m}\gamma_{\alpha,n}^\dagger=\delta_{n,m}\delta_{\alpha,\alpha'}$ and  
$\gamma_{\alpha,n}\gamma_{\alpha',m}+\gamma_{\alpha',m}\gamma_{\alpha,n}=0$ for $\alpha,\alpha'=1,2,3,4$. Then we can plug Eq. (\ref{eq:tr1}) into Eq. (\ref{eq:OP0}) and obtain 
\begin{equation} \label{eq:OP1}
\begin{split}
\braket{c_{r,k,\sigma}c_{r,-k,\bar\sigma}}  =
\sum_{n}u_{n} v_{n}^* 
\braket{1-\gamma_{\alpha,n}^\dagger\gamma_{\alpha,n} -\gamma_{\beta,n}^\dagger\gamma_{\beta,n} }
\end{split}
\end{equation}
where we used commutation relation of $\gamma$ and we have suppressed $r,k,\ud$ index in right-hand side of Eq. (\ref{eq:OP1}) for brevity.
For finite temperature, $\braket{\gamma_{\alpha,n}^\dagger\gamma_{\beta,m}}=\delta_{n,m}\delta_{\alpha,\beta} f(E_n)$ and $\braket{\gamma_{\alpha,n}\gamma_{\beta,m}}=0$, where $f(E)$ is Fermi-Dirac distribution. Therefore, we obtain the s-wave pairing order parameter in Eq. (\ref{eq:OP0}) 
\begin{equation} \label{eq:OP2}
\begin{split}
\Psi(r,k)=&\braket{c_{r,k,\uparrow}c_{r,-k,\downarrow}} \\
=&\sum_{n}u_{n,r,k} v_{n,r,k}^*  (1-2f(E_{n,r,k}))  \\
=& \sum_{n}u_{n,r,k} v_{n,r,k}^*  \tanh \frac{E_{n,r,k}}{2k_BT},
\end{split}
\end{equation}
and the resultant mean-field pairing Hamiltonian is then
\begin{equation}
H_{int}=\sum_{r,k} \Delta(r,k)c^\dagger_{r,k,\uparrow}c^\dagger_{r,-k,\downarrow}
 +h.c.
\end{equation} 
where $\Delta(r,k) = g\Psi(r,k) $ and $g>0$ is an attractive interaction strength for the order paramter definition of Eq. (\ref{eq:OP0}).

\pagebreak
\clearpage
\onecolumngrid

\setcounter{equation}{0}
\setcounter{figure}{0}
\setcounter{table}{0}
\setcounter{page}{1}
\makeatletter
\renewcommand{\theequation}{S\arabic{equation}}
\renewcommand{\thefigure}{S\arabic{figure}}
\renewcommand{\bibnumfmt}[1]{[S#1]}
\renewcommand{\citenumfont}[1]{S#1}
\begin{center}
\textbf{\large Supplementary on ``Probing unconventional superconductivity in inversion symmetric doped Weyl semimetal''}
\end{center}

In the main text, we assume that a supercurrent, $J_S$, is well below the critical current and is proportional to a center-of-mass momentum, $q$, of the Cooper pair. Based on this assumption, our main results are presented as a function of $q$ instead of $J_S$. To see the linear dependency of the current on $q$, we compute a current from an expectation value of a single particle current operator. We utilize the ground states of the general BdG Hamiltonian whose center of momentum is shifted by $q$ and develop formalisms to compute a current as a function of momentum $q$. At the end of this supplement, we use normal BCS Hamiltonian and doped Weyl semimetal Hamiltonian to show the current-momentum ($q$) relationship that follows our assumption; the current is proportional to $q$ before it reaches $q_c$ after which the current reaches the maximum and decreases.

We consider number operator $\hat{N}_{n,\sigma}=c_{n,\sigma}^\dagger c_{n,\sigma}$ and a hopping Hamiltonian $\hat{H}_{hop,\sigma}=\hat{H}_{n+1,n,\sigma}+\hat{H}_{n,n+1,\sigma}=t (c^\dagger_{n,\sigma} c_{n+1,\sigma}+h.c.)$ in transport direction, where $\sigma=\uparrow,\downarrow$ is spin index, $n$ is a site index in transport direction, and $t$ is a hopping constant. The left moving mass flow operator can be defined at cite $n$ as
\begin{equation} \label{eq:c1}
\frac{d\hat{N}_{n,\sigma}}{dt}=\frac{1}{i\hbar}[\hat{N}_{n,\sigma},\hat{H}_{hop,\sigma}]=\frac{t}{i\hbar}[c_{n,\sigma}^\dagger c_{n,\sigma}, c_{n,\sigma'}^\dagger c_{n+1,\sigma'} +  c_{n+1,\sigma'}^\dagger c_{n,\sigma'}].
\end{equation}
By summing over the spatial and spin space and using ground state eigenvectors, we have total current
\begin{equation} \label{eq:c2}
\begin{split}
I_{\sigma}(q)=&
\left\langle \sum_{n,\sigma'}  \frac{d\hat{N}_{n,\sigma}}{dt}  \right\rangle
=\frac{t}{i\hbar} \bra{\Omega_q}
\sum_{n,\sigma'}[c_{n,\sigma}^\dagger c_{n,\sigma}, c_{n,\sigma'}^\dagger c_{n+1,\sigma'}+c_{n+1,\sigma'}^\dagger c_{n+1,\sigma'}] \ket{\Omega_q}, \\
\end{split}
\end{equation}
where $x_n=an$ with lattice constant $a$ and $\ket{\Omega_q}$ represents ground state of the system at $q$. 
The Eq. (\ref{eq:c2}) is Fourier transformed to 
\begin{equation} \label{eq:c3}
\begin{split}
&I_{\sigma}(q)=\frac{t}{i\hbar}\sum_{n,\sigma'} \bra{\Omega_q}
[c_{n,\sigma}^\dagger c_{n,\sigma}, c_{n,\sigma'}^\dagger c_{n+1,\sigma'}+c_{n+1,\sigma'}^\dagger c_{n+1,\sigma'}] \ket{\Omega_q} \\
=&  
\sum_{n,\sigma'} \int \frac{dk_1}{2\pi} \frac{dk_2}{2\pi}\frac{dk_3}{2\pi}\frac{dk_4}{2\pi}
I_{\sigma\sigma'}(k_1,k_2,k_3,k_4)
e^{-i(k_1-k_2+k_3-k_4)x_n}  \\
=&   
\sum_{\sigma'} \int \frac{dk_1}{2\pi} \frac{dk_2}{2\pi}\frac{dk_3}{2\pi}\frac{dk_4}{2\pi}
I_{\sigma\sigma'}(k_1,k_2,k_3,k_4)
2\pi\delta(k_1-k_2+k_3-k_4),  \\
\end{split}
\end{equation} 
where $k_{1,2,3,4}$ are integral variables in momentum space, and we define 
\begin{equation} \label{eq:Iqk}
\begin{split}
I_{\sigma\sigma'}(k_1,k_2,k_3,k_4)=&\frac{t}{i\hbar}
\langle[c^\dagger_{k_1,\sigma} c_{k_2,\sigma}
,c_{k_3,\sigma'}^\dagger c_{k_4,\sigma'}e^{ik_4 a}+c_{k_3,\sigma'}^\dagger c_{k_4,\sigma'}e^{-ik_3 a}]\rangle. \\
=&\frac{t}{i\hbar}
\langle[c^\dagger_{k_1,\sigma} c_{k_2,\sigma}
,c_{k_3,\sigma'}^\dagger c_{k_4,\sigma'}e^{ik_4 a}\rangle
+\frac{t}{i\hbar}
\langle[c^\dagger_{k_1,\sigma} c_{k_2,\sigma}
,c_{k_3,\sigma'}^\dagger c_{k_4,\sigma'}e^{-ik_3 a}]\rangle. \\
\end{split}
\end{equation}
By the Wick's theorem, $\langle c_1 c_2 c_3 c_4 \rangle_0=\langle c_1 c_2 \rangle_0\langle c_3 c_4\rangle_0 + \langle c_1 c_4 \rangle_0\langle c_2 c_3\rangle_0 - \langle c_1 c_3 \rangle_0\langle c_2 c_4\rangle_0$ where a minus sign is from odd number of fermionic operator permutation. Then the first term in right-hand side of Eq. (\ref{eq:Iqk}) is
\begin{equation}
\langle
c^\dagger_1 c_2 c^\dagger_3 c_4 - c^\dagger_3 c_4 c^\dagger_1 c_2
\rangle e^{ik_4 a}
=\left ( \langle c^\dagger_1 c_4 \rangle \langle c_2 c_3^\dagger \rangle 
-
\langle c^\dagger_3 c_2 \rangle \langle c_4 c_1^\dagger \rangle
-
\langle c^\dagger_1 c_3^\dagger \rangle \langle c_2 c_4 \rangle 
+
\langle c^\dagger_3 c_1^\dagger \rangle \langle c_4 c_2 \rangle \right)  e^{ik_4 a}.
\end{equation}
where we simplified the notation $c_{k_{i},\sigma}\rightarrow c_{i}$ for $i=1,2$ and $c_{k_{j},\sigma'}\rightarrow c_{j}$ for $j=3,4$. Similarily, the second term in right-hand side of Eq. (\ref{eq:Iqk}) is
\begin{equation} \label{eq:ck3}
\begin{split}
\langle
c^\dagger_1 c_2 c^\dagger_3 c_4 - c^\dagger_3 c_4 c^\dagger_1 c_2
\rangle e^{-ik_3 a}
=&\left( \langle c^\dagger_1 c_4 \rangle \langle c_2 c_3^\dagger \rangle 
-
\langle c^\dagger_3 c_2 \rangle \langle c_4 c_1^\dagger \rangle 
-
\langle c^\dagger_1 c_3^\dagger \rangle \langle c_2 c_4 \rangle 
+
\langle c^\dagger_3 c_1^\dagger \rangle \langle c_4 c_2 \rangle \right) e^{-ik_3 a} \\
=&
-\left( 
\langle
c^\dagger_1 c_2 c^\dagger_3 c_4 - c^\dagger_3 c_4 c^\dagger_1 c_2
\rangle e^{ik_4 a}
\right)^\dagger,
\end{split}
\end{equation}
where we exchange $3\leftrightarrow4$ and $1\leftrightarrow2$ in second line of Eq. (\ref{eq:ck3}) without losing generality due to the fact that $k_{1,2,3,4}$ are integral variables.
As a result, Eq. (\ref{eq:Iqk}) is simplified as
\begin{equation} \label{eq:Iqk2}
\begin{split}
I_{\sigma\sigma'}(k_1,k_2,k_3,k_4)=& \frac{t}{i\hbar}\left(
\langle
c^\dagger_1 c_2 c^\dagger_3 c_4 - c^\dagger_3 c_4 c^\dagger_1 c_2
\rangle e^{ik_4 a}
-( 
\langle
c^\dagger_1 c_2 c^\dagger_3 c_4 - c^\dagger_3 c_4 c^\dagger_1 c_2
\rangle e^{ik_4 a}
)^\dagger \right) \\
=& \frac{2t}{\hbar}\text{Im}\left\{ 
\left ( \langle c^\dagger_1 c_4 \rangle \langle c_2 c_3^\dagger \rangle 
-
\langle c^\dagger_3 c_2 \rangle \langle c_4 c_1^\dagger \rangle
-
\langle c^\dagger_1 c_3^\dagger \rangle \langle c_2 c_4 \rangle 
+
\langle c^\dagger_3 c_1^\dagger \rangle \langle c_4 c_2 \rangle \right)  e^{ik_4 a}
\right\},
\end{split}
\end{equation}
where $\text{Im}\{A\}$ takes an imaginary part of $A$. As a result, we obtain current in momentum space in single-particle picture. In order to compute Eq. (\ref{eq:Iqk2}), it is useful to transform creation and annihilation operators of the electron ($c^\dagger_k$, $c_k$) to quasi-particle operators ($\gamma_k^\dagger$, $\gamma_k$). For a given Hamiltonian $\mathbf{H}$ whose matrix form is Hermitian is diagonalized as $\mathbf{HR=RE}$, where $\mathbf{E}$ is a diagonal matrix containing eigenvalues of the system and $\mathbf{R}$ is a rotational matrix which maps electron operators to quasi-particle operators. For example, for a arbitrary spinor containing $N$ basis, $\mathbf\Psi=[c_{1}, c_{2},c_{3},\dots,c_{N}]^T$, electron operators are transformed to quasi-particle operators $\mathbf\Gamma=[\gamma_{1},\gamma_{2},\gamma_{3},\dots,\gamma_{N}]^T$ by taking $\mathbf{\Psi=R\Gamma}$. Namely, 

\begin{equation} \label{eq:cRgamma}
\begin{pmatrix}
c_{1}\\ c_{2}\\ \vdots \\ c_{N}
\end{pmatrix}
=
\begin{pmatrix}
(\dotsm R_{1i} \dotsm ) \\
(\dotsm R_{2i} \dotsm ) \\
\vdots \\
(\dotsm R_{Ni} \dotsm ) \\
\end{pmatrix}
\begin{pmatrix}
\gamma_{1}\\ \gamma_{2}\\ \vdots \\ \gamma_{N} \\ 
\end{pmatrix}
\text{ or } c_{l}=\sum_{i}^N R_{li} \gamma_{i},
\end{equation}
where $R_{ij}$ is ($i,j$) element of matrix $\mathbf{R}$.
In our particular case, the Nambu spinor is given as $\Psi=[c_{k+q,\uparrow}, c_{k+q,\downarrow}$ $,c_{-k+q,\uparrow}^\dagger,c_{-k+q,\downarrow}^\dagger]^T$ whose center-of-mass momentum is shifted by $q$. Then the electron creation and annihilation operators are
\begin{equation}
c_{k+q,\uparrow}=\sum_{i} R_{1i} \gamma_{k,i}, \;
c_{k+q,\downarrow}=\sum_{i} R_{2i} \gamma_{k,i}, \;
c_{-k+q,\uparrow}^\dagger=\sum_{i} R_{3i} \gamma_{k,i}, \;
c_{-k+q,\downarrow}^\dagger=\sum_{i} R_{4i} \gamma_{k,i},
\end{equation}
where quasi-particle operator is defined as $\mathbf{\Gamma}_k=[\gamma_{k,1},\gamma_{k,2},\gamma_{k,3},\gamma_{k,4}]^T$.
Note that quasi-particle operator satisfies $\gamma^\dagger_i\gamma_j+\gamma_j\gamma_i^\dagger=\delta_{ij}$, $\langle \gamma_i^\dagger \gamma_j \rangle=\delta_{ij}f(E_i)$, $\langle \gamma_i \gamma_j^\dagger \rangle=\delta_{ij}(1-f(E_i))$, and $\langle \gamma_i\gamma_j \rangle=\langle \gamma_i^\dagger\gamma_j^\dagger \rangle=0$, where $f(E)$ is a Fermi-dirac function. We now compute Eq. (\ref{eq:Iqk2}). For example,
\begin{equation}
\begin{split}
\langle c^\dagger_{k_1,\uparrow} c_{k_4,\uparrow} \rangle e^{ik_4 a}
=&
\langle c^\dagger_{k+q,\uparrow} c_{k+q,\uparrow} \rangle e^{i(k+q) a}
+\langle c^\dagger_{k+q,\uparrow} c_{-k+q,\uparrow} \rangle e^{i(-k+q) a}
+\langle c^\dagger_{-k+q,\uparrow} c_{k+q,\uparrow} \rangle e^{i(k+q) a}
+\langle c^\dagger_{-k+q,\uparrow} c_{-k+q,\uparrow} \rangle e^{i(-k+q) a} \\
=&
\langle \sum_{i,j} R^\dagger_{1i} \gamma^\dagger_{k,i} R_{1j} \gamma_{k,j} \rangle e^{i(k+q) a}
+\langle \sum_{i,j} R^\dagger_{1i} \gamma^\dagger_{k,i} R_{3j}^\dagger \gamma^\dagger_{k,j} \rangle e^{i(-k+q) a} \\
+&
\langle \sum_{i,j} R_{3i} \gamma_{k,i} R_{1j} \gamma_{k,j} \rangle e^{i(k+q) a}
+\langle \sum_{i,j} R_{3i} \gamma_{k,i} R^\dagger_{3j} \gamma^\dagger_{k,j}  \rangle e^{i(-k+q) a} \\
=&
 \sum_i R^\dagger_{1i}R_{1i} \langle \gamma^\dagger_{k,i}\gamma_{k,i} \rangle e^{i(k+q) a}
+ \sum_i R^\dagger_{3i}R_{3i} \langle \gamma_{k,i}\gamma^\dagger_{k,i} \rangle e^{i(-k+q) a} \\
=&\left(
\sum_i R^\dagger_{1i}R_{1i} f(E_{i,k}) 
\right)e^{i(k+q) a}
+
\left(
\sum_i R^\dagger_{3i}R_{3i}[1-f(E_{i,k})] 
\right)e^{i(-k+q) a}.
\end{split}
\end{equation}
Similar calculation is carried out for all possible permutations of spin and basis combinations. By defining 
\begin{equation} \label{eq:F}
\begin{split}
F_{mn}^{e}=&\sum_i R_{m i}^\dagger R_{n i} f(E_{i,k}) ,\;\;
F_{mn}^{h}=\sum_i R_{m i}^\dagger R_{n i}[1-f(E_{i,k})],\;\;
(F_{mn}^{e})^\dagger = F_{nm}^{e},\;\; (F_{mn}^{h})^\dagger = F_{nm}^{h},
\end{split}
\end{equation}
the first term in second line of Eq. (\ref{eq:Iqk2}) is
\begin{equation} \label{eq:cop3_1}
\begin{split}
\langle c^\dagger_{1,\sigma} c_{4,\sigma'} \rangle \langle c_{2,\sigma} c_{3,\sigma'}^\dagger \rangle e^{ik_4 a}  
=&\left( 
\langle c^\dagger_{1,\uparrow} c_{4,\uparrow} \rangle \langle c_{2,\uparrow} c_{3,\uparrow}^\dagger \rangle 
+
\langle c^\dagger_{1,\downarrow} c_{4,\downarrow} \rangle \langle c_{2,\downarrow} c_{3,\downarrow}^\dagger \rangle  
+ 
\langle c^\dagger_{1,\uparrow} c_{4,\downarrow} \rangle \langle c_{2,\uparrow} c_{3,\downarrow}^\dagger \rangle 
+
\langle c^\dagger_{1,\downarrow} c_{4,\uparrow} \rangle \langle c_{2,\downarrow} c_{3,\uparrow}^\dagger \rangle  
\right)e^{ik_4 a} \\
=&\left( F_{11}^{h} + F_{33}^{e} \right)_1 \left( F_{11}^{e} e^{ika} + F_{33}^{h} e^{-ika} \right)_2 e^{iqa}
+\left( F_{22}^{h} + F_{44}^{e} \right)_1 \left( F_{22}^{e} e^{ika} + F_{44}^{h} e^{-ika} \right)_2 e^{iqa} \\
+&\left( F_{21}^{h} + F_{34}^{e} \right)_1 \left( F_{12}^{e} e^{ika} + F_{43}^{h} e^{-ika} \right)_2 e^{iqa} 
+\left( F_{12}^{h} + F_{43}^{e} \right)_1 \left( F_{21}^{e} e^{ika} + F_{34}^{h} e^{-ika} \right)_2 e^{iqa}, \\
\end{split}
\end{equation}
where a subscript $(\dotsm)_{1,2}$ is an integration variable index. The second term in second line of Eq. (\ref{eq:Iqk2}) is
\begin{equation} \label{eq:cop3_2}
\begin{split}
\langle c^\dagger_{3,\sigma'} c_{2,\sigma} \rangle \langle c_{4,\sigma'} c_{1,\sigma}^\dagger \rangle e^{ik_4 a}
=&\left( 
\langle c^\dagger_{3,\uparrow} c_{2,\uparrow} \rangle \langle c_{4,\uparrow} c_{1,\uparrow}^\dagger \rangle 
+
\langle c^\dagger_{3,\downarrow} c_{2,\downarrow} \rangle \langle c_{4,\downarrow} c_{1,\downarrow}^\dagger \rangle  
+ 
\langle c^\dagger_{3,\uparrow} c_{2,\downarrow} \rangle \langle c_{4,\uparrow} c_{1,\downarrow}^\dagger \rangle 
+
\langle c^\dagger_{3,\downarrow} c_{2,\uparrow} \rangle \langle c_{4,\downarrow} c_{1,\uparrow}^\dagger \rangle  
\right)e^{ik_4 a} \\
=&\left( F_{11}^{h} e^{ika} + F_{33}^{e} e^{-ika} \right)_1 \left( F_{11}^{e}  + F_{33}^{h}  \right)_2 e^{iqa}
+\left( F_{22}^{h} e^{ika} + F_{44}^{e} e^{-ika} \right)_1 \left( F_{22}^{e}  + F_{44}^{h}  \right)_2 e^{iqa} \\
+&\left( F_{21}^{h} e^{ika} + F_{34}^{e} e^{-ika} \right)_1 \left( F_{12}^{e}  + F_{43}^{h}  \right)_2 e^{iqa}
+\left( F_{12}^{h} e^{ika} + F_{43}^{e} e^{-ika} \right)_1 \left( F_{21}^{e}  + F_{34}^{h}  \right)_2 e^{iqa}. \\
\end{split}
\end{equation}
The third term in right hand side of Eq. (\ref{eq:Iqk2}) is
\begin{equation} \label{eq:cop3_3}
\begin{split}
\langle c^\dagger_{1,\sigma} c_{3,\sigma'}^\dagger \rangle \langle c_{2,\sigma} c_{4,\sigma'} \rangle e^{ik_4 a}
=&\left( 
\langle c^\dagger_{1,\uparrow} c_{3,\uparrow}^\dagger \rangle \langle c_{2,\uparrow} c_{4,\uparrow} \rangle 
+
\langle c^\dagger_{1,\downarrow} c_{3,\downarrow}^\dagger \rangle \langle c_{2,\downarrow} c_{4,\downarrow} \rangle   
+ 
\langle c^\dagger_{1,\uparrow} c_{3,\downarrow}^\dagger \rangle \langle c_{2,\uparrow} c_{4,\downarrow} \rangle 
+
\langle c^\dagger_{1,\downarrow} c_{3,\uparrow}^\dagger \rangle \langle c_{2,\downarrow} c_{4,\uparrow} \rangle 
\right)e^{ik_4 a} \\
=&\left( F_{13}^{h} + F_{13}^{e} \right)_1 \left( F_{31}^{e} e^{ika} + F_{31}^{h} e^{-ika} \right)_2 e^{iqa} 
+\left( F_{24}^{h} + F_{24}^{e} \right)_1 \left( F_{42}^{e} e^{ika} + F_{42}^{h} e^{-ika} \right)_2 e^{iqa} \\
+&\left( F_{23}^{h} + F_{14}^{e} \right)_1 \left( F_{32}^{e} e^{ika} + F_{41}^{h} e^{-ika} \right)_2 e^{iqa} 
+\left( F_{23}^{e} + F_{14}^{h} \right)_1 \left( F_{41}^{e} e^{ika} + F_{32}^{h} e^{-ika} \right)_2 e^{iqa}. \\
\end{split}
\end{equation}
The fourth term in right hand side of Eq. (\ref{eq:Iqk2}) is
\begin{equation} \label{eq:cop3_4}
\begin{split}
\langle c^\dagger_{3,\sigma'} c_{1,\sigma}^\dagger \rangle \langle c_{4,\sigma'} c_{2,\sigma} \rangle e^{ik_4 a}
=&\left( 
\langle c^\dagger_{3,\uparrow} c_{1,\uparrow}^\dagger \rangle \langle c_{4,\uparrow} c_{2,\uparrow} \rangle 
+
\langle c^\dagger_{3,\downarrow} c_{1,\downarrow}^\dagger \rangle \langle c_{4,\downarrow} c_{2,\downarrow} \rangle   
+ 
\langle c^\dagger_{3,\uparrow} c_{1,\downarrow}^\dagger \rangle \langle c_{4,\uparrow} c_{2,\downarrow} \rangle 
+
\langle c^\dagger_{3,\downarrow} c_{1,\uparrow}^\dagger \rangle \langle c_{4,\downarrow} c_{2,\uparrow} \rangle 
\right)e^{ik_4 a} \\
=&\left( F_{13}^{h} + F_{13}^{e} \right)_1 \left( F_{31}^{e} e^{-ika} + F_{31}^{h} e^{ika} \right)_2 e^{iqa} 
+\left( F_{24}^{h} + F_{24}^{e} \right)_1 \left( F_{42}^{e} e^{-ika} + F_{42}^{h} e^{ika} \right)_2 e^{iqa} \\
+&\left( F_{23}^{h} + F_{14}^{e} \right)_1 \left( F_{32}^{e} e^{-ika} + F_{41}^{h} e^{ika} \right)_2 e^{iqa} 
+\left( F_{23}^{e} + F_{14}^{h} \right)_1 \left( F_{41}^{e} e^{-ika} + F_{32}^{h} e^{ika} \right)_2 e^{iqa}. \\
\end{split}
\end{equation}
Therefore, after we diagonalize the Hamiltonian and obtain rotational matrix $\mathbf{R}$, the single particle current, $I(q)$ in Eq. (\ref{eq:c3}), is computed by plugging Eqs. (\ref{eq:cop3_1}-\ref{eq:cop3_4}) into Eq. (\ref{eq:Iqk2}) and integrating over momentum space. This procedure is valid for arbitrary Hamiltonian. In case of a typical metallic Hamiltonian with simple parabolic dispersion, the rotational matrix is\cite{SGennes1966}
\begin{equation}
R=
\begin{pmatrix}
u & 0 &  0 & -v \\
0 & u & -v & 0 \\
0 & v &  u & 0 \\
v & 0 &  0 & u \\
\end{pmatrix}
\text{with} \;
HR=R
\begin{pmatrix}
-E_k & 0 &  0 & 0 \\
0 & -E_k & 0 & 0 \\
0 & 0 & E_k & 0 \\
0 & 0 &  0 & E_k \\
\end{pmatrix}
\end{equation}
where $u^2+v^2=1$. Then we have non-zero component for $R_{ii}$, $R_{14}$, $R_{41}$, $R_{23}$, and $R_{32}$ only. At zero temperature, Eq. (\ref{eq:F}) results in
\begin{equation}
\begin{split}
&F_{11}^{e}=u^2,\; F_{11}^{h}=v^2, \;
F_{22}^{e}=u^2,\; F_{22}^{h}=v^2 , \\
&F_{33}^{e}=v^2,\; F_{33}^{h}=u^2, \;
F_{44}^{e}=v^2,\; F_{44}^{h}=u^2 , \\
&F_{14}^{e}=uv,\; F_{14}^{h}=-uv, \;
F_{23}^{e}=uv,\; F_{23}^{h}=-uv.
\end{split}
\end{equation}
As a result, Eqs. (\ref{eq:cop3_1}), (\ref{eq:cop3_2}, (\ref{eq:cop3_3}, (\ref{eq:cop3_4}) are
\begin{equation} \label{eq:cop3_5}
\begin{split}
\langle c^\dagger_{1,\sigma} c_{4,\sigma'} \rangle \langle c_{2,\sigma} c_{3,\sigma'}^\dagger \rangle e^{ik_4 a}  
=&8v_1^2 u_2^2 \cos(k_2 a) e^{iqa} \\
\langle c^\dagger_{3,\sigma'} c_{2,\sigma} \rangle \langle c_{4,\sigma'} c_{1,\sigma}^\dagger \rangle e^{ik_4 a}
=& 8u_1^2 v_2^2 \cos(k_2 a) e^{iqa} \\
\langle c^\dagger_{1,\sigma} c_{3,\sigma'}^\dagger \rangle \langle c_{2,\sigma} c_{4,\sigma'} \rangle e^{ik_4 a}
=& 
\langle c^\dagger_{3,\sigma'} c_{1,\sigma}^\dagger \rangle \langle c_{4,\sigma'} c_{2,\sigma} \rangle e^{ik_4 a}
=0.
\end{split}
\end{equation}
Consequently, the single particle current is 
\begin{equation} \label{eq:Iqmetal}
\begin{split}
I(q)=&\frac{t}{i\hbar}\langle[c^\dagger_{n,\sigma} c_{n,\sigma},
c_{n,\sigma'}^\dagger c_{n+1,\sigma'}+c_{n+1,\sigma'}^\dagger c_{n,\sigma'}]\rangle \\
=&
\frac{8t}{\hbar}\sin(qa)\int\int\frac{dk_1}{2\pi}\frac{dk_2}{2\pi} (v_1^2 u_2^2-u_1^2 v_2^2) \cos(k_2 a).
\end{split}
\end{equation}

\begin{figure}[h!]  
  \centering
   \includegraphics[width=0.5\textwidth]{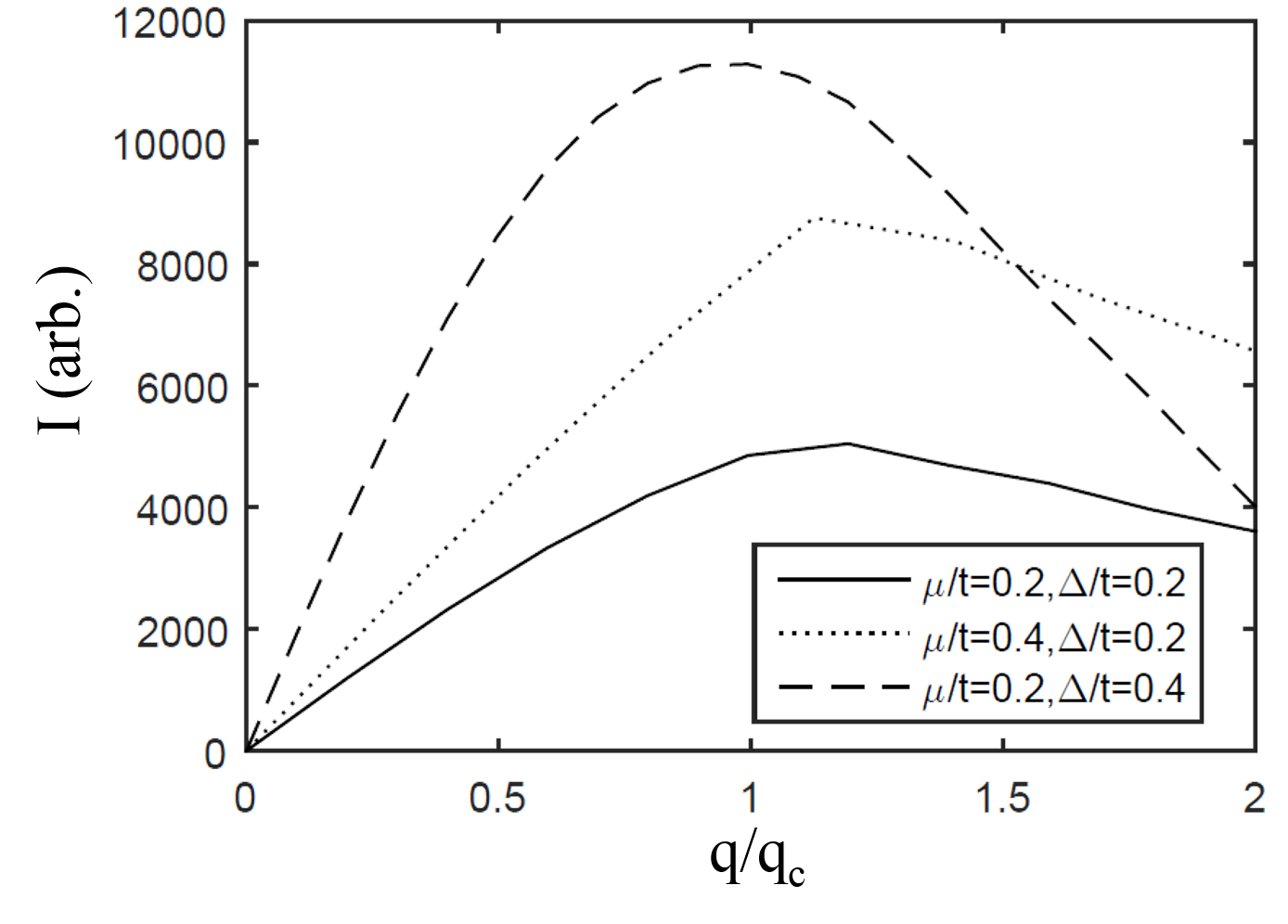}
  \caption{A plot of uniform supercurrent calculated for a metallic Hamiltonian with parabolic band.}\label{fig:I_BCS}
\end{figure}

In case of a normal BCS superconductor, Eq. (\ref{eq:Iqmetal}) shows that the current is proportional to $q$ when $qa\ll1$ as $\sin(qa)\simeq qa$. Specifically, the s-wave superconductor BdG Hamiltonian in Eq. (\ref{eq:HRzq}) is diagonalized and we obtain $\mathbf{R}$ of Eq. (\ref{eq:cRgamma}). Then we calculate current from Eq. (\ref{eq:c3}) using Eqs. (\ref{eq:cop3_1}-\ref{eq:cop3_4}). Fig. \ref{fig:I_BCS} shows the resultant current as a function of momentum $q$. The current is linear in $q$ until the current reaches the maximum. 
We note that the maximum current in Fig. \ref{fig:I_BCS} corresponds to the critical current\cite{SSauls1995}.
Fig. \ref{fig:I_BCS} is qualitatively explained by following. A uniform supercurrent which shifts a center-of-momentum frame of the Cooper pair by $q$ introduces the Doppler shift in energy spectrum $E(\mathbf{k},\mathbf{q}) = E_0(\mathbf{k}) + \mathbf{p}\cdot \mathbf{v}_s$ where $E_0(\mathbf{k})$ is energy spectrum in equilibrium and $\mathbf{v}_s=(\hbar/m^*)\mathbf{q}$ is an effective velocity of the Cooper pair center-of-mass frame\cite{SBardeen1962, SGennes1966}. When the velocity reaches \emph{dephasing velocity}, $v_c=\Delta/\hbar k_F$, or $q$ reaches $q_c=m^*\Delta/\hbar^2 k_F$, the excitation gap of the superconducting system is closed at $k=-k_F$. In case of $q>q_c$, the quasi-particles are populated even at zero temperature and, consequently, the superconducting phase becomes unstable\cite{SBardeen1962, SGennes1966}.  

\begin{figure}[h!]  
  \centering
   \includegraphics[width=0.5\textwidth]{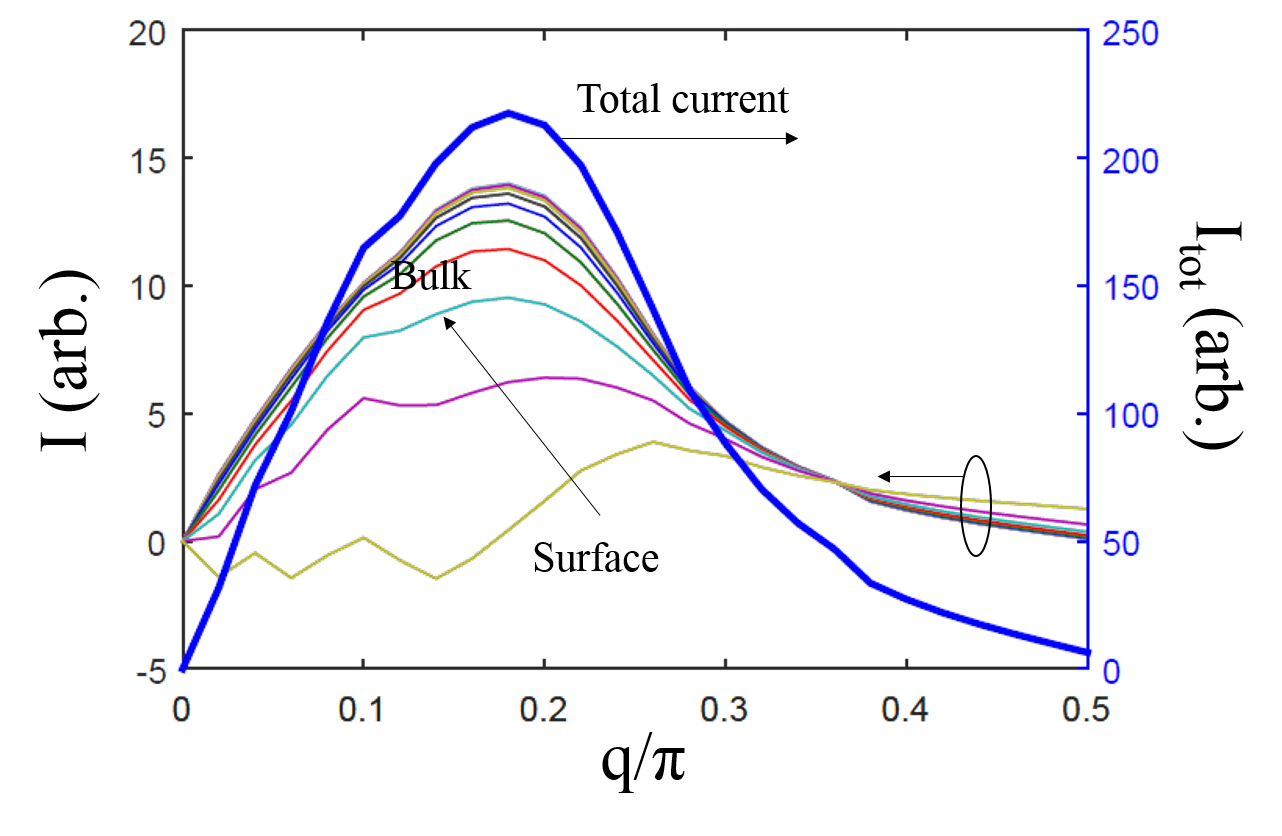}
  \caption{A plot of uniform supercurrent calculated for a Weyl semimetal Hamiltonian. $\hat{x},\hat{z}$ direction is periodic and $\hat{y}$ direction is open with the thickness of $N_y=20$. The current is applied in $\hat{z}$ direction parallel to the axis where the nodal points are located. Left vertical axis is for spatial resolved current $I(y)$ and right vertical axis is for total current.}\label{fig:I_Weyl}
\end{figure}

Similarly, the BdG Hamiltonian in Eq. (\ref{eq:HBdGw}) is constructed using a doped Weyl semimetal Hamiltonian and a current is calculated from Eq. (\ref{eq:c3}) using Eqs. (\ref{eq:cop3_1}-\ref{eq:cop3_4}). The system is discretized in thickness ($\hat{y}$) direction ($N_y=20$, periodic boundary condition in $\hat{x}$, $\hat{z}$ directions) in order to observe the surface states contribution on the current. Fig. \ref{fig:I_Weyl} shows the resultant current as a function of momentum $q$. When $q$ is small, we observe a current as a function of $q$ whose trend is in a close analogy with the supercurrent of the nodal superconductors\cite{SMaki2004, SMaki2004_2}. Although the system has the Fermi arc, we find that the finite density of states at surface has minimal contribution in current in our particular case. The ``topological flow'', defined as $\mathbf{v}_{topo}=\mathbf{n}\times\mathbf{v}$ where $\mathbf{n}$ is a surface normal vector ($\hat{y}$) and $\mathbf{v}$ is a group velocity of the surface states, flows along the axis where the nodal points are aligned due to the fact that the topological flow flows from a nodal point of positive vorticity to another nodal point of negative vorticity\cite{SMurakami2014, SSato2015, SHaldane2015}. If the current direction is parallel to the topological flow ($\mathbf{q}\parallel\mathbf{v}_{topo}$ and, thus, $\mathbf{q}\bot\mathbf{v}$), which is in $\hat{z}$ direction in our case, the group velocity of the Fermi arc in current direction is zero. The resultant surface state contribution on the total current is small as it is observed in Fig \ref{fig:I_Weyl}. In contrast, we observe a finite amount of surface current when we choose a direction of a uniform current perpendicular to the topological flow ($\mathbf{q}\bot\mathbf{v}_{topo}$ and, thus, $\mathbf{q}\parallel\mathbf{v}$) due to the fact that the Fermi arc has a net group velocity in current direction (not shown here). 
With minimal contribution of the surface states on the total current, the current response of the nodal BCS states in doped WSM may show similar results with that of the nodal superconductors. Indeed, Fig. \ref{fig:I_Weyl} shows a current that is proportional to $q$ before it reaches $q_c$ after which the current becomes the maximum, which is similarily observed in calculated current of $d$ and $f$ wave superconductors\cite{SMaki2004, SMaki2004_2}. 

\begin{thebibliography}{31}
\expandafter\ifx\csname natexlab\endcsname\relax\def\natexlab#1{#1}\fi
\expandafter\ifx\csname bibnamefont\endcsname\relax
  \def\bibnamefont#1{#1}\fi
\expandafter\ifx\csname bibfnamefont\endcsname\relax
  \def\bibfnamefont#1{#1}\fi
\expandafter\ifx\csname citenamefont\endcsname\relax
  \def\citenamefont#1{#1}\fi
\expandafter\ifx\csname url\endcsname\relax
  \def\url#1{\texttt{#1}}\fi
\expandafter\ifx\csname urlprefix\endcsname\relax\def\urlprefix{URL }\fi
\providecommand{\bibinfo}[2]{#2}
\providecommand{\eprint}[2][]{\url{#2}}

\bibitem[{\citenamefont{Murakami}(2007)}]{Murakami2007}
\bibinfo{author}{\bibfnamefont{S.}~\bibnamefont{Murakami}},
  \bibinfo{journal}{New Journal of Physics} \textbf{\bibinfo{volume}{9}},
  \bibinfo{pages}{356} (\bibinfo{year}{2007}),
  \urlprefix\url{http://stacks.iop.org/1367-2630/9/i=9/a=356}.

\bibitem[{\citenamefont{Wan et~al.}(2011)\citenamefont{Wan, Turner, Vishwanath,
  and Savrasov}}]{Savrasov2011}
\bibinfo{author}{\bibfnamefont{X.}~\bibnamefont{Wan}},
  \bibinfo{author}{\bibfnamefont{A.~M.} \bibnamefont{Turner}},
  \bibinfo{author}{\bibfnamefont{A.}~\bibnamefont{Vishwanath}},
  \bibnamefont{and} \bibinfo{author}{\bibfnamefont{S.~Y.}
  \bibnamefont{Savrasov}}, \bibinfo{journal}{Phys. Rev. B}
  \textbf{\bibinfo{volume}{83}}, \bibinfo{pages}{205101}
  (\bibinfo{year}{2011}),
  \urlprefix\url{http://link.aps.org/doi/10.1103/PhysRevB.83.205101}.

\bibitem[{\citenamefont{Chiu and Schnyder}(2014)}]{Schnyder2014}
\bibinfo{author}{\bibfnamefont{C.-K.} \bibnamefont{Chiu}} \bibnamefont{and}
  \bibinfo{author}{\bibfnamefont{A.~P.} \bibnamefont{Schnyder}},
  \bibinfo{journal}{Phys. Rev. B} \textbf{\bibinfo{volume}{90}},
  \bibinfo{pages}{205136} (\bibinfo{year}{2014}),
  \urlprefix\url{http://link.aps.org/doi/10.1103/PhysRevB.90.205136}.

\bibitem[{\citenamefont{Fang et~al.}(2003)\citenamefont{Fang, Nagaosa,
  Takahashi, Asamitsu, Mathieu, Ogasawara, Yamada, Kawasaki, Tokura, and
  Terakura}}]{Fang2003}
\bibinfo{author}{\bibfnamefont{Z.}~\bibnamefont{Fang}},
  \bibinfo{author}{\bibfnamefont{N.}~\bibnamefont{Nagaosa}},
  \bibinfo{author}{\bibfnamefont{K.~S.} \bibnamefont{Takahashi}},
  \bibinfo{author}{\bibfnamefont{A.}~\bibnamefont{Asamitsu}},
  \bibinfo{author}{\bibfnamefont{R.}~\bibnamefont{Mathieu}},
  \bibinfo{author}{\bibfnamefont{T.}~\bibnamefont{Ogasawara}},
  \bibinfo{author}{\bibfnamefont{H.}~\bibnamefont{Yamada}},
  \bibinfo{author}{\bibfnamefont{M.}~\bibnamefont{Kawasaki}},
  \bibinfo{author}{\bibfnamefont{Y.}~\bibnamefont{Tokura}}, \bibnamefont{and}
  \bibinfo{author}{\bibfnamefont{K.}~\bibnamefont{Terakura}},
  \bibinfo{journal}{Science} \textbf{\bibinfo{volume}{302}},
  \bibinfo{pages}{92} (\bibinfo{year}{2003}),
  \urlprefix\url{http://www.sciencemag.org/content/302/5642/92.abstract}.

\bibitem[{\citenamefont{Nielsen and
  Ninomiya}(1981{\natexlab{a}})}]{Nielsen1981_1}
\bibinfo{author}{\bibfnamefont{H.}~\bibnamefont{Nielsen}} \bibnamefont{and}
  \bibinfo{author}{\bibfnamefont{M.}~\bibnamefont{Ninomiya}},
  \bibinfo{journal}{Nuclear Physics B} \textbf{\bibinfo{volume}{185}},
  \bibinfo{pages}{20 } (\bibinfo{year}{1981}{\natexlab{a}}), ISSN
  \bibinfo{issn}{0550-3213},
  \urlprefix\url{http://www.sciencedirect.com/science/article/pii/0550321381903618}.

\bibitem[{\citenamefont{Nielsen and
  Ninomiya}(1981{\natexlab{b}})}]{Nielsen1981_2}
\bibinfo{author}{\bibfnamefont{H.}~\bibnamefont{Nielsen}} \bibnamefont{and}
  \bibinfo{author}{\bibfnamefont{M.}~\bibnamefont{Ninomiya}},
  \bibinfo{journal}{Nuclear Physics B} \textbf{\bibinfo{volume}{193}},
  \bibinfo{pages}{173 } (\bibinfo{year}{1981}{\natexlab{b}}), ISSN
  \bibinfo{issn}{0550-3213},
  \urlprefix\url{http://www.sciencedirect.com/science/article/pii/0550321381905241}.

\bibitem[{\citenamefont{Okugawa and Murakami}(2014)}]{Murakami2014}
\bibinfo{author}{\bibfnamefont{R.}~\bibnamefont{Okugawa}} \bibnamefont{and}
  \bibinfo{author}{\bibfnamefont{S.}~\bibnamefont{Murakami}},
  \bibinfo{journal}{Phys. Rev. B} \textbf{\bibinfo{volume}{89}},
  \bibinfo{pages}{235315} (\bibinfo{year}{2014}),
  \urlprefix\url{http://link.aps.org/doi/10.1103/PhysRevB.89.235315}.

\bibitem[{\citenamefont{Xu et~al.}(2015)\citenamefont{Xu, Belopolski, Alidoust,
  Neupane, Bian, Zhang, Sankar, Chang, Yuan, Lee et~al.}}]{Xu2015}
\bibinfo{author}{\bibfnamefont{S.-Y.} \bibnamefont{Xu}},
  \bibinfo{author}{\bibfnamefont{I.}~\bibnamefont{Belopolski}},
  \bibinfo{author}{\bibfnamefont{N.}~\bibnamefont{Alidoust}},
  \bibinfo{author}{\bibfnamefont{M.}~\bibnamefont{Neupane}},
  \bibinfo{author}{\bibfnamefont{G.}~\bibnamefont{Bian}},
  \bibinfo{author}{\bibfnamefont{C.}~\bibnamefont{Zhang}},
  \bibinfo{author}{\bibfnamefont{R.}~\bibnamefont{Sankar}},
  \bibinfo{author}{\bibfnamefont{G.}~\bibnamefont{Chang}},
  \bibinfo{author}{\bibfnamefont{Z.}~\bibnamefont{Yuan}},
  \bibinfo{author}{\bibfnamefont{C.-C.} \bibnamefont{Lee}},
  \bibnamefont{et~al.}, \bibinfo{journal}{Science}
  \textbf{\bibinfo{volume}{349}}, \bibinfo{pages}{613} (\bibinfo{year}{2015}),
  \urlprefix\url{http://www.sciencemag.org/content/349/6248/613.abstract}.

\bibitem[{\citenamefont{Hosur and Qi}(2013)}]{Hosur2013}
\bibinfo{author}{\bibfnamefont{P.}~\bibnamefont{Hosur}} \bibnamefont{and}
  \bibinfo{author}{\bibfnamefont{X.}~\bibnamefont{Qi}},
  \bibinfo{journal}{Comptes Rendus Physique} \textbf{\bibinfo{volume}{14}},
  \bibinfo{pages}{857 } (\bibinfo{year}{2013}), ISSN \bibinfo{issn}{1631-0705},
  \bibinfo{note}{topological insulators / Isolants topologiquesTopological
  insulators / Isolants topologiques},
  \urlprefix\url{http://www.sciencedirect.com/science/article/pii/S1631070513001710}.

\bibitem[{\citenamefont{Cho et~al.}(2012)\citenamefont{Cho, Bardarson, Lu, and
  Moore}}]{Cho2012}
\bibinfo{author}{\bibfnamefont{G.~Y.} \bibnamefont{Cho}},
  \bibinfo{author}{\bibfnamefont{J.~H.} \bibnamefont{Bardarson}},
  \bibinfo{author}{\bibfnamefont{Y.-M.} \bibnamefont{Lu}}, \bibnamefont{and}
  \bibinfo{author}{\bibfnamefont{J.~E.} \bibnamefont{Moore}},
  \bibinfo{journal}{Phys. Rev. B} \textbf{\bibinfo{volume}{86}},
  \bibinfo{pages}{214514} (\bibinfo{year}{2012}),
  \urlprefix\url{http://link.aps.org/doi/10.1103/PhysRevB.86.214514}.

\bibitem[{\citenamefont{Wei et~al.}(2014)\citenamefont{Wei, Chao, and
  Aji}}]{Vivek2014}
\bibinfo{author}{\bibfnamefont{H.}~\bibnamefont{Wei}},
  \bibinfo{author}{\bibfnamefont{S.-P.} \bibnamefont{Chao}}, \bibnamefont{and}
  \bibinfo{author}{\bibfnamefont{V.}~\bibnamefont{Aji}},
  \bibinfo{journal}{Phys. Rev. B} \textbf{\bibinfo{volume}{89}},
  \bibinfo{pages}{014506} (\bibinfo{year}{2014}),
  \urlprefix\url{http://link.aps.org/doi/10.1103/PhysRevB.89.014506}.

\bibitem[{\citenamefont{Bednik et~al.}(2015)\citenamefont{Bednik, Zyuzin, and
  Burkov}}]{Burkov2015}
\bibinfo{author}{\bibfnamefont{G.}~\bibnamefont{Bednik}},
  \bibinfo{author}{\bibfnamefont{A.~A.} \bibnamefont{Zyuzin}},
  \bibnamefont{and} \bibinfo{author}{\bibfnamefont{A.~A.}
  \bibnamefont{Burkov}}, \bibinfo{journal}{Phys. Rev. B}
  \textbf{\bibinfo{volume}{92}}, \bibinfo{pages}{035153}
  (\bibinfo{year}{2015}),
  \urlprefix\url{http://link.aps.org/doi/10.1103/PhysRevB.92.035153}.

\bibitem[{\citenamefont{Zhou et~al.}(2016)\citenamefont{Zhou, Gao, and
  Wang}}]{Wang2015}
\bibinfo{author}{\bibfnamefont{T.}~\bibnamefont{Zhou}},
  \bibinfo{author}{\bibfnamefont{Y.}~\bibnamefont{Gao}}, \bibnamefont{and}
  \bibinfo{author}{\bibfnamefont{Z.~D.} \bibnamefont{Wang}},
  \bibinfo{journal}{Phys. Rev. B} \textbf{\bibinfo{volume}{93}},
  \bibinfo{pages}{094517} (\bibinfo{year}{2016}),
  \urlprefix\url{http://link.aps.org/doi/10.1103/PhysRevB.93.094517}.

\bibitem[{\citenamefont{Hosur et~al.}(2014)\citenamefont{Hosur, Dai, Fang, and
  Qi}}]{Qi2014}
\bibinfo{author}{\bibfnamefont{P.}~\bibnamefont{Hosur}},
  \bibinfo{author}{\bibfnamefont{X.}~\bibnamefont{Dai}},
  \bibinfo{author}{\bibfnamefont{Z.}~\bibnamefont{Fang}}, \bibnamefont{and}
  \bibinfo{author}{\bibfnamefont{X.-L.} \bibnamefont{Qi}},
  \bibinfo{journal}{Phys. Rev. B} \textbf{\bibinfo{volume}{90}},
  \bibinfo{pages}{045130} (\bibinfo{year}{2014}),
  \urlprefix\url{http://link.aps.org/doi/10.1103/PhysRevB.90.045130}.

\bibitem[{\citenamefont{Lu et~al.}(2015)\citenamefont{Lu, Yada, Sato, and
  Tanaka}}]{Sato2015}
\bibinfo{author}{\bibfnamefont{B.}~\bibnamefont{Lu}},
  \bibinfo{author}{\bibfnamefont{K.}~\bibnamefont{Yada}},
  \bibinfo{author}{\bibfnamefont{M.}~\bibnamefont{Sato}}, \bibnamefont{and}
  \bibinfo{author}{\bibfnamefont{Y.}~\bibnamefont{Tanaka}},
  \bibinfo{journal}{Phys. Rev. Lett.} \textbf{\bibinfo{volume}{114}},
  \bibinfo{pages}{096804} (\bibinfo{year}{2015}),
  \urlprefix\url{http://link.aps.org/doi/10.1103/PhysRevLett.114.096804}.

\bibitem[{\citenamefont{Li and Haldane}()}]{Haldane2015}
\bibinfo{author}{\bibfnamefont{Y.}~\bibnamefont{Li}} \bibnamefont{and}
  \bibinfo{author}{\bibfnamefont{F.~D.~M.} \bibnamefont{Haldane}},
  \bibinfo{journal}{arXiv:1510.01730}  (????).

\bibitem[{\citenamefont{Fulde and Ferrell}(1964)}]{Ferrell1964}
\bibinfo{author}{\bibfnamefont{P.}~\bibnamefont{Fulde}} \bibnamefont{and}
  \bibinfo{author}{\bibfnamefont{R.~A.} \bibnamefont{Ferrell}},
  \bibinfo{journal}{Phys. Rev.} \textbf{\bibinfo{volume}{135}},
  \bibinfo{pages}{A550} (\bibinfo{year}{1964}),
  \urlprefix\url{http://link.aps.org/doi/10.1103/PhysRev.135.A550}.

\bibitem[{\citenamefont{Larkin and Ovchinnikov}(1965)}]{Ovchinnikov1965}
\bibinfo{author}{\bibfnamefont{A.}~\bibnamefont{Larkin}} \bibnamefont{and}
  \bibinfo{author}{\bibfnamefont{Y.}~\bibnamefont{Ovchinnikov}},
  \bibinfo{journal}{Sov. Phys. JETP} \textbf{\bibinfo{volume}{20}},
  \bibinfo{pages}{762} (\bibinfo{year}{1965}).

\bibitem[{\citenamefont{Yang and Agterberg}(2000)}]{Yang2000}
\bibinfo{author}{\bibfnamefont{K.}~\bibnamefont{Yang}} \bibnamefont{and}
  \bibinfo{author}{\bibfnamefont{D.~F.} \bibnamefont{Agterberg}},
  \bibinfo{journal}{Phys. Rev. Lett.} \textbf{\bibinfo{volume}{84}},
  \bibinfo{pages}{4970} (\bibinfo{year}{2000}),
  \urlprefix\url{http://link.aps.org/doi/10.1103/PhysRevLett.84.4970}.

\bibitem[{\citenamefont{Ashby and Carbotte}(2013)}]{Carbotte2013}
\bibinfo{author}{\bibfnamefont{P.~E.~C.} \bibnamefont{Ashby}} \bibnamefont{and}
  \bibinfo{author}{\bibfnamefont{J.~P.} \bibnamefont{Carbotte}},
  \bibinfo{journal}{Phys. Rev. B} \textbf{\bibinfo{volume}{87}},
  \bibinfo{pages}{245131} (\bibinfo{year}{2013}),
  \urlprefix\url{http://link.aps.org/doi/10.1103/PhysRevB.87.245131}.

\bibitem[{\citenamefont{Gennes}(1966)}]{Gennes1966}
\bibinfo{author}{\bibfnamefont{P.~D.} \bibnamefont{Gennes}},
  \emph{\bibinfo{title}{Superconductivity of Metals and Alloys}}
  (\bibinfo{publisher}{W. A. Benjamin, New York}, \bibinfo{year}{1966}).

\bibitem[{\citenamefont{Tinkham}(1996)}]{Tinkham1996}
\bibinfo{author}{\bibfnamefont{M.}~\bibnamefont{Tinkham}},
  \emph{\bibinfo{title}{Introduction to Superconductivity}}
  (\bibinfo{publisher}{McGraw-Hill Inc.}, \bibinfo{year}{1996}).

\bibitem[{\citenamefont{Bardeen}(1962)}]{Bardeen1962}
\bibinfo{author}{\bibfnamefont{J.}~\bibnamefont{Bardeen}},
  \bibinfo{journal}{Rev. Mod. Phys.} \textbf{\bibinfo{volume}{34}},
  \bibinfo{pages}{667} (\bibinfo{year}{1962}),
  \urlprefix\url{http://link.aps.org/doi/10.1103/RevModPhys.34.667}.

\bibitem[{\citenamefont{Khavkine et~al.}(2004)\citenamefont{Khavkine, Kee, and
  Maki}}]{Maki2004}
\bibinfo{author}{\bibfnamefont{I.}~\bibnamefont{Khavkine}},
  \bibinfo{author}{\bibfnamefont{H.-Y.} \bibnamefont{Kee}}, \bibnamefont{and}
  \bibinfo{author}{\bibfnamefont{K.}~\bibnamefont{Maki}},
  \bibinfo{journal}{Phys. Rev. B} \textbf{\bibinfo{volume}{70}},
  \bibinfo{pages}{184521} (\bibinfo{year}{2004}),
  \urlprefix\url{http://link.aps.org/doi/10.1103/PhysRevB.70.184521}.

\bibitem[{\citenamefont{Yang et~al.}(2011)\citenamefont{Yang, Lu, and
  Ran}}]{Ran2011}
\bibinfo{author}{\bibfnamefont{K.-Y.} \bibnamefont{Yang}},
  \bibinfo{author}{\bibfnamefont{Y.-M.} \bibnamefont{Lu}}, \bibnamefont{and}
  \bibinfo{author}{\bibfnamefont{Y.}~\bibnamefont{Ran}},
  \bibinfo{journal}{Phys. Rev. B} \textbf{\bibinfo{volume}{84}},
  \bibinfo{pages}{075129} (\bibinfo{year}{2011}),
  \urlprefix\url{http://link.aps.org/doi/10.1103/PhysRevB.84.075129}.

\bibitem[{\citenamefont{Shapourian and Hughes}(2016)}]{Hughes2016}
\bibinfo{author}{\bibfnamefont{H.}~\bibnamefont{Shapourian}} \bibnamefont{and}
  \bibinfo{author}{\bibfnamefont{T.~L.} \bibnamefont{Hughes}},
  \bibinfo{journal}{Phys. Rev. B} \textbf{\bibinfo{volume}{93}},
  \bibinfo{pages}{075108} (\bibinfo{year}{2016}),
  \urlprefix\url{http://link.aps.org/doi/10.1103/PhysRevB.93.075108}.

\bibitem[{\citenamefont{Hal\'asz and Balents}(2012)}]{Balants2012}
\bibinfo{author}{\bibfnamefont{G.~B.} \bibnamefont{Hal\'asz}} \bibnamefont{and}
  \bibinfo{author}{\bibfnamefont{L.}~\bibnamefont{Balents}},
  \bibinfo{journal}{Phys. Rev. B} \textbf{\bibinfo{volume}{85}},
  \bibinfo{pages}{035103} (\bibinfo{year}{2012}),
  \urlprefix\url{http://link.aps.org/doi/10.1103/PhysRevB.85.035103}.

\bibitem[{\citenamefont{Chen and Franz}()}]{Franz2016}
\bibinfo{author}{\bibfnamefont{A.}~\bibnamefont{Chen}} \bibnamefont{and}
  \bibinfo{author}{\bibfnamefont{M.}~\bibnamefont{Franz}},
  \bibinfo{journal}{arXiv:1601.01727}  (????).

\bibitem[{\citenamefont{Datta}(2005)}]{Datta2005}
\bibinfo{author}{\bibfnamefont{S.}~\bibnamefont{Datta}},
  \emph{\bibinfo{title}{Quantum Transport: Atom to Transistor}},
  科学前沿丛书 (\bibinfo{publisher}{Cambridge University Press},
  \bibinfo{year}{2005}), ISBN \bibinfo{isbn}{9780521631457},
  \urlprefix\url{https://books.google.com/books?id=Yj50EJoS224C}.

\bibitem[{\citenamefont{Xu et~al.}(1995)\citenamefont{Xu, Yip, and
  Sauls}}]{Sauls1995}
\bibinfo{author}{\bibfnamefont{D.}~\bibnamefont{Xu}},
  \bibinfo{author}{\bibfnamefont{S.~K.} \bibnamefont{Yip}}, \bibnamefont{and}
  \bibinfo{author}{\bibfnamefont{J.~A.} \bibnamefont{Sauls}},
  \bibinfo{journal}{Phys. Rev. B} \textbf{\bibinfo{volume}{51}},
  \bibinfo{pages}{16233} (\bibinfo{year}{1995}),
  \urlprefix\url{http://link.aps.org/doi/10.1103/PhysRevB.51.16233}.

\bibitem[{\citenamefont{Kee et~al.}(2004)\citenamefont{Kee, Kim, and
  Maki}}]{Maki2004_2}
\bibinfo{author}{\bibfnamefont{H.-Y.} \bibnamefont{Kee}},
  \bibinfo{author}{\bibfnamefont{Y.~B.} \bibnamefont{Kim}}, \bibnamefont{and}
  \bibinfo{author}{\bibfnamefont{K.}~\bibnamefont{Maki}},
  \bibinfo{journal}{Phys. Rev. B} \textbf{\bibinfo{volume}{70}},
  \bibinfo{pages}{052505} (\bibinfo{year}{2004}),
  \urlprefix\url{http://link.aps.org/doi/10.1103/PhysRevB.70.052505}.

\end{thebibliography}

\begin{thebibliography}{8}%
\makeatletter
\providecommand \@ifxundefined [1]{%
 \@ifx{#1\undefined}
}%
\providecommand \@ifnum [1]{%
 \ifnum #1\expandafter \@firstoftwo
 \else \expandafter \@secondoftwo
 \fi
}%
\providecommand \@ifx [1]{%
 \ifx #1\expandafter \@firstoftwo
 \else \expandafter \@secondoftwo
 \fi
}%
\providecommand \natexlab [1]{#1}%
\providecommand \enquote  [1]{``#1''}%
\providecommand \bibnamefont  [1]{#1}%
\providecommand \bibfnamefont [1]{#1}%
\providecommand \citenamefont [1]{#1}%
\providecommand \href@noop [0]{\@secondoftwo}%
\providecommand \href [0]{\begingroup \@sanitize@url \@href}%
\providecommand \@href[1]{\@@startlink{#1}\@@href}%
\providecommand \@@href[1]{\endgroup#1\@@endlink}%
\providecommand \@sanitize@url [0]{\catcode `\\12\catcode `\$12\catcode
  `\&12\catcode `\#12\catcode `\^12\catcode `\_12\catcode `\%12\relax}%
\providecommand \@@startlink[1]{}%
\providecommand \@@endlink[0]{}%
\providecommand \url  [0]{\begingroup\@sanitize@url \@url }%
\providecommand \@url [1]{\endgroup\@href {#1}{\urlprefix }}%
\providecommand \urlprefix  [0]{URL }%
\providecommand \Eprint [0]{\href }%
\providecommand \doibase [0]{http://dx.doi.org/}%
\providecommand \selectlanguage [0]{\@gobble}%
\providecommand \bibinfo  [0]{\@secondoftwo}%
\providecommand \bibfield  [0]{\@secondoftwo}%
\providecommand \translation [1]{[#1]}%
\providecommand \BibitemOpen [0]{}%
\providecommand \bibitemStop [0]{}%
\providecommand \bibitemNoStop [0]{.\EOS\space}%
\providecommand \EOS [0]{\spacefactor3000\relax}%
\providecommand \BibitemShut  [1]{\csname bibitem#1\endcsname}%
\let\auto@bib@innerbib\@empty
\bibitem [{\citenamefont {Gennes}(1966)}]{SGennes1966}%
  \BibitemOpen
  \bibfield  {author} {\bibinfo {author} {\bibfnamefont {P.~D.}\ \bibnamefont
  {Gennes}},\ }\href@noop {} {\emph {\bibinfo {title} {Superconductivity of
  Metals and Alloys}}}\ (\bibinfo  {publisher} {W. A. Benjamin, New York},\
  \bibinfo {year} {1966})\BibitemShut {NoStop}%
\bibitem [{\citenamefont {Xu}\ \emph {et~al.}(1995)\citenamefont {Xu},
  \citenamefont {Yip},\ and\ \citenamefont {Sauls}}]{SSauls1995}%
  \BibitemOpen
  \bibfield  {author} {\bibinfo {author} {\bibfnamefont {D.}~\bibnamefont
  {Xu}}, \bibinfo {author} {\bibfnamefont {S.~K.}\ \bibnamefont {Yip}}, \ and\
  \bibinfo {author} {\bibfnamefont {J.~A.}\ \bibnamefont {Sauls}},\ }\href
  {\doibase 10.1103/PhysRevB.51.16233} {\bibfield  {journal} {\bibinfo
  {journal} {Phys. Rev. B}\ }\textbf {\bibinfo {volume} {51}},\ \bibinfo
  {pages} {16233} (\bibinfo {year} {1995})}\BibitemShut {NoStop}%
\bibitem [{\citenamefont {Bardeen}(1962)}]{SBardeen1962}%
  \BibitemOpen
  \bibfield  {author} {\bibinfo {author} {\bibfnamefont {J.}~\bibnamefont
  {Bardeen}},\ }\href {\doibase 10.1103/RevModPhys.34.667} {\bibfield
  {journal} {\bibinfo  {journal} {Rev. Mod. Phys.}\ }\textbf {\bibinfo {volume}
  {34}},\ \bibinfo {pages} {667} (\bibinfo {year} {1962})}\BibitemShut
  {NoStop}%
\bibitem [{\citenamefont {Khavkine}\ \emph {et~al.}(2004)\citenamefont
  {Khavkine}, \citenamefont {Kee},\ and\ \citenamefont {Maki}}]{SMaki2004}%
  \BibitemOpen
  \bibfield  {author} {\bibinfo {author} {\bibfnamefont {I.}~\bibnamefont
  {Khavkine}}, \bibinfo {author} {\bibfnamefont {H.-Y.}\ \bibnamefont {Kee}}, \
  and\ \bibinfo {author} {\bibfnamefont {K.}~\bibnamefont {Maki}},\ }\href
  {\doibase 10.1103/PhysRevB.70.184521} {\bibfield  {journal} {\bibinfo
  {journal} {Phys. Rev. B}\ }\textbf {\bibinfo {volume} {70}},\ \bibinfo
  {pages} {184521} (\bibinfo {year} {2004})}\BibitemShut {NoStop}%
\bibitem [{\citenamefont {Kee}\ \emph {et~al.}(2004)\citenamefont {Kee},
  \citenamefont {Kim},\ and\ \citenamefont {Maki}}]{SMaki2004_2}%
  \BibitemOpen
  \bibfield  {author} {\bibinfo {author} {\bibfnamefont {H.-Y.}\ \bibnamefont
  {Kee}}, \bibinfo {author} {\bibfnamefont {Y.~B.}\ \bibnamefont {Kim}}, \ and\
  \bibinfo {author} {\bibfnamefont {K.}~\bibnamefont {Maki}},\ }\href {\doibase
  10.1103/PhysRevB.70.052505} {\bibfield  {journal} {\bibinfo  {journal} {Phys.
  Rev. B}\ }\textbf {\bibinfo {volume} {70}},\ \bibinfo {pages} {052505}
  (\bibinfo {year} {2004})}\BibitemShut {NoStop}%
\bibitem [{\citenamefont {Okugawa}\ and\ \citenamefont
  {Murakami}(2014)}]{SMurakami2014}%
  \BibitemOpen
  \bibfield  {author} {\bibinfo {author} {\bibfnamefont {R.}~\bibnamefont
  {Okugawa}}\ and\ \bibinfo {author} {\bibfnamefont {S.}~\bibnamefont
  {Murakami}},\ }\href {\doibase 10.1103/PhysRevB.89.235315} {\bibfield
  {journal} {\bibinfo  {journal} {Phys. Rev. B}\ }\textbf {\bibinfo {volume}
  {89}},\ \bibinfo {pages} {235315} (\bibinfo {year} {2014})}\BibitemShut
  {NoStop}%
\bibitem [{\citenamefont {Lu}\ \emph {et~al.}(2015)\citenamefont {Lu},
  \citenamefont {Yada}, \citenamefont {Sato},\ and\ \citenamefont
  {Tanaka}}]{SSato2015}%
  \BibitemOpen
  \bibfield  {author} {\bibinfo {author} {\bibfnamefont {B.}~\bibnamefont
  {Lu}}, \bibinfo {author} {\bibfnamefont {K.}~\bibnamefont {Yada}}, \bibinfo
  {author} {\bibfnamefont {M.}~\bibnamefont {Sato}}, \ and\ \bibinfo {author}
  {\bibfnamefont {Y.}~\bibnamefont {Tanaka}},\ }\href {\doibase
  10.1103/PhysRevLett.114.096804} {\bibfield  {journal} {\bibinfo  {journal}
  {Phys. Rev. Lett.}\ }\textbf {\bibinfo {volume} {114}},\ \bibinfo {pages}
  {096804} (\bibinfo {year} {2015})}\BibitemShut {NoStop}%
\bibitem [{\citenamefont {Li}\ and\ \citenamefont {Haldane}()}]{SHaldane2015}%
  \BibitemOpen
  \bibfield  {author} {\bibinfo {author} {\bibfnamefont {Y.}~\bibnamefont
  {Li}}\ and\ \bibinfo {author} {\bibfnamefont {F.~D.~M.}\ \bibnamefont
  {Haldane}},\ }\href@noop {} {\bibinfo  {journal} {arXiv:1510.01730}\
  }\BibitemShut {NoStop}%
\end{thebibliography}
\end{document}